\documentstyle[12pt]{article}
\begin{document}
\setlength{\oddsidemargin}{1.0in}
\setlength{\evensidemargin}{1.0in}
\baselineskip 0.55cm
\title{
Rotating metrics admitting non-perfect fluids in General
Relativity}

\author
{Ng. Ibohal \\
Department of Mathematics, University of Manipur\\
Imphal-795003, Manipur, India\\
E-mail: (i) ngibohal@rediffmail.com\\
(ii) ngibohal@iucaa.ernet.in}
\date{\today}
\maketitle

\begin{abstract}
         In this paper an application of
Newman-Janis algorithm in spherical symmetric metrics with the
functions $M(u,r)$ and $e(u,r)$ has been discussed. After the
transformation of the metric via this algorithm, these two
functions $M(u,r)$ and $e(u,r)$ will be of the three variables
$u,r,\theta$. With these functions of three variables, all the
Newman-Penrose (NP) spin coefficients, the Ricci as well as the
Weyl scalars have been calculated from the Cartan's structure
equations. From these NP quantities, a class of rotating solutions
of Einstein's field equations can be obtained. These solutions
include (a) Kerr-Newman, (b) rotating Vaidya solution, (c)
rotating Vaidya-Bonnor solution, (d) rotating Husain's solution,
(e) rotating Wang-Wu solutions. It is found that the technique
developed by Wang and Wu can be used to generate new embedded
solutions, that the Kerr-Newman solution can be combined smoothly
with the rotating Vaidya solution to generate Kerr-Newman-Vaidya
solution, and similarly, Kerr-Newman-Vaidya-Bonnor solution of the
field equations. It has also shown that the embedded universes
like Kerr-Newman de Sitter, rotating Vaidya-Bonnor-de Sitter,
Kerr-Newman-Vaidya-de Sitter can be derived from the general
solutions with Wang-Wu function. All rotating embedded solutions
derived here can be written in Kerr-Schild forms, showing the
extension of Xanthopoulos's theorem. It is also found that all the
rotating solutions admit non-perfect fluids.
\end{abstract}

\vspace{.25in}

\section{Introduction}

\setcounter{equation}{0}

\renewcommand{\theequation}{\arabic{equation}}

In an earlier paper [1] it is shown that Hawking's radiation [2]
could be expressed in classical spacetime metrics, by considering
the charge $e$ to be function of the radial coordinate $r$ of
Reissner-Nordstrom as well as Kerr-Newman black holes. Since these
two black holes describe the `stationary' metrics, it has intended
to search for `non-stationary' rotating metrics in order to
incorporate relativistic aspect of Hawking's radiation in general
relativity. The non-rotating Vaidya metric is a non-stationary
generalization of Schwarzschild vacuum solution, describing the
gravitational field of a null radiating star. Many attempts have
been made to generate non-stationary rotating metrics which
describe the rotating external gravitational field of radiating
bodies. To mention with, Vaidya and Patel [3] obtained a
non-stationary rotating metric with mass $M=-m\{1
+u(b/a^2)-(b^2/4a^2)\cos^2\theta\}^{3/2}$ having minus sign, where
$m$, $a$, $b$ are constant and $u$ is the retarded time
coordinate. They claimed that their metric recovers the Kerr
metric when $b=0$. However, it is well known that the Kerr metric
has mass without negative sign. Carmeli and Kaye [4] have shown,
by considering the mass $M$ of Kerr metric directly as function of
coordinate $u$, that the Kerr metric can be made a non-stationary
rotating metric, called variable mass Kerr metric. However,
Herrera and Martinez [5], Herrera et. al. [6] have discussed the
interpretation of variable mass Kerr metric of Carmeli and Kaye.
Gonzalez et.al.[7] have presented a non-stationary generalization
of the Kerr-Newman metric, by allowing the three parameters $a$,
$m$ and $e$ to be functions of coordinate $u$ and shown that the
variable $a(u)$ does not represent the rotating electromagnetic
field of the Einstein-Maxwell equations and concluded that to take
the parameters of a metric as functions of $u$ does not generalize
the solutions enough. Mallett [8] applied Newman-Janis algorithm
to the Reissner-Nordstrom-de Sitter `seed' solution to derive a
Kerr-Newman-de Sitter solution and afterward he considered the
mass and the charged to be the functions of the retarded time
coordinate $u$ in order to get non-stationary charged radiating
metric. Xu [9] has discussed the nature of the field equations of
Mallett's solution. Jing and Wang [10] have considered the mass
$M(u)$ and the charge $e(u)$ unchanged after the application of
Newman-Janis algorithm to the non-rotating Vaidya-Bonnor `seed'
solution with mass $M(u)$ and charge $e(u)$. In fact, after the
application of the algorithm, the mass $M(u)$ and charge $e(u)$
should be functions $M(u,\theta)$, $e(u,\theta)$ of two variables
$u$ and $\theta$.

Here we employ the Newman-Janis algorithm [11] to generate
rotating non-stationary metrics from the spherically symmetric
`seed' metric with the functions $M(u,r)$ and $e(u,r)$, where $u$
and $r$ are the coordinates of the spacetime geometry.
Newman-Janis algorithm [11] is a {\it complex coordinate
transformation}, which has been introduced by Newman and Janis to
obtain Kerr metric, a rotating Schwarzschild {\sl vacuum} solution
of Einstein's field equations from the non-rotating Schwarzschild
`seed' solution. In another paper Newman et.al. [12] again applied
the same transformation to the non-rotating charged
Reissner-Nordstrom solution to get rotating charged
Reissner-Nordstrom solution, which is now commonly known as
Kerr-Newman black hole solution in General Relativity. So this
complex coordinate transformation can be used to derive rotating
solutions from the non-rotating `seed' solutions of Einstein's
equations of spherical symmetric metrics. Herrera and Jimenez [13]
applied the same Newman-Janis algorithm to an interior
non-rotating spherically symmetric seed metric and the resulting
rotating interior was tried to match with the exterior Kerr metric
on the boundary of the source. Drake and Turolla [14] have
generated a class of metrics as possible sources for the Kerr
metric by applying the same algorithm to any static spherically
symmetric `seed' metric. Drake and Szekeres [15] have shown the
uniqueness of this algorithm in generating the Kerr-Newman metric
and proved that the only electrovac Petrov type D spacetime
generated by the algorithm with a vanishing Ricci scalar $\Lambda$
is the Kerr-Newman space-time. Yazadjiev [16] has also shown that
Sen's rotating dilation-axiom black-hole solution [17] can be
derived from the static spherically symmetric dilation black hole
solution via this algorithm too.

The purposes of this paper are
\begin{enumerate}
\item to apply the Newman-Janis algorithm to the spherical
symmetric `seed' metric with the functions $M(u,r)$ and $e(u,r)$
of two variables $u$, $r$, \item to calculate all the
Newman-Penrose (NP) spin coefficients, the Ricci as well as the
Weyl scalars [18] in general, \item to give examples of rotating
solutions, published and unpublished, of Einstein's field
equations from these NP quantities.
\end{enumerate}
The spherically symmetric metric with the functions of two
variables $u$, $r$ has been transformed via Newman-Janis algorithm
[11] to get rotating metrics. After the transformation, $M$ and
$e$ will be functions of three variables $u,r,\theta$. Then we
calculate all the Newman-Penrose (NP) spin coefficients, the Ricci
as well as the Weyl scalars in general. Accordingly, the
Einstein's tensors as well as the energy momentum tensors (EMT) of
the matter fields have been presented in terms of complex null
tetrad vectors. From this  EMT one observes the description of
having two fluids system in the field equations. To visualize the
two fluid system we rewrite the Einstein tensors in terms of one
unit time-like and three unit space-like vectors constructed from
the complex null vectors.

Thus, we can generate rotating solutions mentioned in the abstract
above from the Ricci as well as the Weyl scalars of the
transformed metric. Consequently, some of the results are cited
for ready reference in the form of theorems based on
rotating solutions discussed here.

\vspace*{.10in}

\newtheorem{theorem}{Theorem}
\begin{theorem}
If $g_{ab}^{\rm KN}$ is the Kerr-Newman solution of Einstein's
field equations and $\ell_a$ is geodesic, shear free, rotating and
expanding null vector and one of the double repeated principal
null directions of the Weyl tensor of $g_{ab}^{\rm KN}$, then
$g_{ab}^{\rm KNV}=g_{ab}^{\rm KN}+2Q(u,r,\theta)\,\ell_a\,\ell_b$
will be a {\sl rotating} Kerr-Newman-Vaidya solution with
$Q(u,r,\theta)=-r\,f(u)\,R^{-2}$, where $f(u)$ is the mass
function of {\sl rotating} Vaidya solution.
\end{theorem}
\begin{theorem}
If $g_{ab}^{\rm dS}$ is the rotating de Sitter solution of
Einstein's field equations and $\ell_a$ is geodesic, shear free,
rotating and expanding null vector and one of the double repeated
principal null directions of the Weyl tensor of $g_{ab}^{\rm dS}$,
then $g_{ab}^{\rm KNdS}=g_{ab}^{\rm dS}
+2Q(r,\theta)\,\ell_a\,\ell_b$ will be a Kerr-Newman-de Sitter
solution with $Q(r,\theta)=-(r\,m-e^2/2)\,R^{-2}$, where $m$ and
$e$ are constant and represent the mass and the charge of
Kerr-Newman black hole.
\end{theorem}

\begin{theorem}
All rotating stationary spherically symmetric solutions based on
Newman-Janis algorithm are Petrov type $D$, whose one of the
repeated null vectors, $\ell_a$ is geodesic, shear free, expanding
as well as non-zero twist.
\end{theorem}

\begin{theorem}
All rotating non-stationary spherically symmetric solutions,
derivable from the application of Newman-Janis algorithm and
possessing a geodesic, shear free, expanding and rotating null
vector $\ell_a$, are algebraically special in the Petrov
classification.
\end{theorem}
\vspace*{.10in}

The Kerr-Schild ansatz of Theorem 1 can also be written in another
form as $g_{ab}^{\rm KNV}=g_{ab}^{\rm
V}+2Q(r,\theta)\,\ell_a\,\ell_b$ with
$Q(r,\theta)=-(r\,m-e^2/2)\,R^{-2}$, where $m$ and $e$ are
constant and represent the mass and the charge of Kerr-Newman
black hole and $g_{ab}^{\rm V}$ is a rotating Vaidya solution of
Einstein's field equations and $\ell_a$ is geodesic, shear free,
rotating and expanding null vector of $g_{ab}^{\rm V}$. This
ansatz can be interpreted as the Kerr-Newman black hole embedded
into the rotating Vaidya null radiating background, describing
Kerr-Newman-Vaidya black hole. Similarly, theorem 2 states that
Kerr-Newman black hole can also be embedded into the de Sitter
cosmological background, describing Kerr-Newman-de Sitter  black
hole. Its alternative form will be the ansatz $g_{ab}^{\rm
KNdS}=g_{ab}^{\rm KN}+2Q(r,\theta)\,\ell_a\,\ell_b$ with
$Q(r,\theta)=-(\Lambda^*r^4/6)R^{-2}$, where $\Lambda^*$ is the
cosmological constant. The extension of the theorem 2 in
non-stationary version can be stated in the case of rotating
Vaidya-Bonnor-de Sitter solution. Theorems 3 and 4 follow from the
stationary as well as non-stationary rotating solutions to be
discussed in the next sections.

This paper is organized as follows: Section 2 presents a brief
application of Newman-Janis algorithm to a spherically symmetric
`seed' metric with the functions $M(u,r)$ and $e(u,r)$. A general
expressions of NP quantities with $M(u,r,\theta)$ and
$e(u,r,\theta)$ are calculated from the Cartan's structure
equations and cited for further use in section 3. The general
properties of the rotating spherically symmetric metric is
discussed after observing the nature of the NP quantities. Section
4 discusses the energy conditions of the energy momentum tensor
for a time like observer with its four velocity vector. Section 5
gives examples of rotating stationary as well as non-stationary
solutions from the NP quantities. In section 6 we introduce the
Wang-Wu function in the rotating solutions, and it is shown that
the general rotating solutions with Wang-Wu function can be used
to generate rotating new embedded solutions like Kerr-Newman-de
Sitter, Kerr-Newman-Vaidya, Kerr-Newman-Vaidya-de Sitter, rotating
Vaidya-Bonnor-de Sitter. The conclusion of the paper is cited in
section 7 with suggestions and remarks of the solutions discussed
in the earlier sections.

The presentation of this paper is essentially based on the
Newman-Penrose (NP) spin-coefficient formalism [18]. The NP
quantities are calculated from Cartan's structure equations
written in NP formalism by McIntosh and Hickman [19] in
(+,--,--,--) signature.

\section{Newman-Janis algorithm}
\setcounter{equation}{0}
\renewcommand{\theequation}{2.\arabic{equation}}

For application of Newman-Janis algorithm, we start with a
spherical symmetric `seed' metric written in the form
\begin{equation}
ds^2=e^{2\phi}\,du^2+2du\,dr-r^2(d\theta^2
+{\rm sin}^2\theta\,d\phi^2),
\end{equation}
where
$e^{2\phi} = 1-2M(u,r)/r+e^2(u,r)/r^2$ and the coordinate
chosen are $\{x^1,x^2,x^3,x^4\}$ \\
=$\{u,r,\theta,\phi\}$. The u-coordinate is related to the
retarded time in flat space-time. So u-constant surfaces are null
cones open to the future. The r-constant is null coordinate. The
$\theta$ and $\phi$ are usual angle coordinates. The retarded time
coordinate are used to evaluate the radiating (or outgoing) energy
momentum tensor around the astronomical body [9]. Here $M$ and
$e$ are the functions of the retarded time coordinate $u$ and the
radial coordinate $r$. Initially, when $M$, $e$ are constant, this
metric provides the Reissner-Nordstrom solution and
also when both $M$, $e$ are functions of $u$, it becomes the
non-rotating Vaidya-Bonnor solution. The contravariant components
of the metric (2.1) are
\begin{equation}
g^{ab}  = \left( \begin{array}{cccc}
 0  &   1                 &  0               &                0 \\
 1  &   -{e^{2\phi}/r^2}
                          &  0               &                0 \\

 0  &   0                 & -1/r^2           &                0 \\

 0  &   0                 &  0               & -1/r^2{\rm sin}^2\theta
 \end{array} \right),
\,\;a,b = 1,2,3,4.
\end{equation}

These metric components can be expressed in terms of complex null
tetrad [18]
\begin{eqnarray*}
\ell^a=\delta^a_2,\\
\end{eqnarray*}
\begin{equation}
n^a=\delta^a_1-{1\over 2}(1-{2M\over
r}+{e^2\over r^2})\,\delta^a_2,\\
\end{equation}
\begin{eqnarray*}
m^a={1\over\surd 2r}\,(\delta^a_3+{i\over {\rm
sin}\,\theta}\,\delta^a_4)
\end{eqnarray*}
\begin{eqnarray*}
\overline m^a={1\over\surd 2r}\,(\delta^a_3-{i\over {\rm
sin}\,\theta}\,\delta^a_4).
\end{eqnarray*}
Then the metric tensor $g^{ab}$ of the line element (2.1) is
expressed in
these null tetrad vectors as
\begin{equation}
g^{ab}=\ell^a\,n^b+n^a\,\ell^b-m^a\,\overline m^b-\overline m^a\,m^b.
\end{equation}
Here the null vectors $\ell^a$ and $n^a$ are real, and $m^a$ and
$\overline m^a$ are complex conjugates of each other. According to
Newman and Janis [11], one can complexify the coordinate $r$ and
$u$ by the following transformation
\begin{eqnarray}
r = r'-i\,a\,{\rm cos}\,\theta,\: u = u'+i\,a\,{\rm
cos}\,\theta,\:\theta =\theta',\: \phi = \phi'.
\end{eqnarray}
This complexification can only be done by considering $r'$ and
$u'$ real. Thus, from (2.5) we have the relations
\begin{equation}
dr=dr'+i\,a\,{\rm sin}\,\theta\,d\theta,\;\; du=du'-i\,a\,{\rm
sin}\,\theta\,d\theta,\;\; d\theta=d\theta',\,\, {\rm and}\,\,
d\phi=d\phi'.
\end{equation}
These relations can be utilized in (2.1) to obtain the rotating
metrics:
\begin{eqnarray}
d s^2&=&e^{2\phi}\,du^2+2du\,dr+ 2a\,{\rm
sin}^2\theta(1-e^{2\phi})\,du\,d\phi \cr &&-2a\,{\rm
sin}^2\theta\,dr\,d\phi -R^2d\theta^2-\{R^2-a^2\,{\rm
sin}^2\theta\,(e^{2\phi} -2)\}\,{\rm sin}^2\theta\,d\phi^2,
\end{eqnarray}
where $e^{2\phi} = 1-{2rM(u,r,\theta)/R^2}+{e^2(u,r,\theta)/R^2}$
and $R^2 = r^2 + a^2{\rm cos}^2\theta$. All the primes are being
dropped for convenience of notation. According to Newman and Janis
[11], we have also used a suitable substitution $d\theta =-i\,{\rm
sin}\theta\,d\phi$. Then the covariant complex null tetrad vectors
take the forms
\begin{eqnarray*}
\ell_a=\delta^1_a -a\,{\rm sin}^2\theta\,\delta^4_a,
\end{eqnarray*}
\begin{eqnarray*}
n_a={1\over 2}\,H(u,r,\theta)\,\delta^1_a+ \delta^2_a
-{1\over 2}\,H(u,r,\theta)\,a\,{\rm sin}^2\theta\,\delta^4_a,
\end{eqnarray*}
\begin{equation}
m_a=-{1\over\surd 2R}\,\Big\{-ia\,{\rm
sin}\,\theta\,\delta^1_a+R^2\,\delta^3_a +i(r^2+a^2)\,{\rm
sin}\,\theta\,\delta^4_a\Big\},
\end{equation}
\begin{eqnarray*}
\overline m_a=-{1\over\surd 2\overline R}\,\Big\{ia\,{\rm
sin}\,\theta\,\delta^1_a +R^2\,\delta^3_a-i(r^2+a^2)\,{\rm
sin}\,\theta\,\delta^4_a\Big\}.
\end{eqnarray*}
where $R=r+ia\,{\rm cos}\theta$.
 The null tetrad vectors chosen here are
different from those chosen in [12], but are similar to
those given in Chandrasekhar [20]. Now, after the transformation
(2.5), the functions $M$ and $e$ must be of the three variables
$u,r,\theta$, however the old ones had explicitly $u$ and $r$
dependence. That is,
\begin{equation}
H(u,r,\theta) = 1-{2rM(u,r,\theta)\over R^2}+{e^2(u,r,\theta)\over
R^2}+{a^2\,{\rm sin}^2\theta\over R^2}.
\end{equation}

Then the line element has the covariant components of the metric
tensor $g_{ab}$:
\begin{equation}
g_{ab}  = \left( \begin{array}{cccc}
 e^{2\phi}             &   1                &  0   & a\,{\rm
sin}^2\theta(1-e^{2\phi})         \\
 1                     &   0                &  0   &  -a{\rm sin}^2\theta\,   \\
 0                     &   0                & -R^2 &     0                    \\
 a\,{\rm sin}^2\theta\,(1-e^{2\phi}) & -a\,{\rm sin}^2\theta & 0 &
-\{R^2-a^2\,{\rm sin}^2\theta(e^{2\phi} -2)\}{\rm sin}^2\theta
  \end{array} \right)
\end{equation}
$a,b = 1,2,3,4$. This completes the application of Newman-Janis
algorithm to the spherically symmetric `seed' metric (2.1). The
usefulness of this transformed metric (2.7) will be discussed in the
following sections.

\section{NP quantities for the rotating metric}

\setcounter{equation}{0}
\renewcommand{\theequation}{3.\arabic{equation}}

In this section we shall derive the general NP spin coefficients,
the Ricci scalars and the Weyl scalars for the metric (2.7) and
present the general properties of the metric after observing the
conditions of these NP quantities. First, the basis one-form of
the tetrad vectors (2.8) are given below:
\begin{eqnarray*}
\theta^1&\equiv& n_a\,dx^a = {1\over 2}H\,du + dr - {1\over
2}\,a\,H\,{\rm
sin}^2\theta\,d\phi,
\end{eqnarray*}
\begin{eqnarray*}
\theta^2 &\equiv& \ell_a\,dx^a= du - a\,{\rm
sin}^2\theta\,d\phi,
\end{eqnarray*}
\begin{eqnarray}
\theta^3 &\equiv& -\overline m_a\,dx^a= {1\over\surd 2\overline
R}\Big\{ia\,{\rm sin}\,\theta\,du + R^2\,d\theta
-i(r^2+a^2)\,d\phi\Big\},
\end{eqnarray}
\begin{eqnarray*}
\theta^4 &\equiv& -m_a\,dx^a= {1\over\surd 2\,R} \Big\{-ia\,{\rm
sin}\,\theta\,du +R^2\,d\theta +i(r^2+a^2)\,d\phi\Big\}.
\end{eqnarray*}
The intrinsic derivative operators for the metric (2.7) take the
following forms:
\begin{eqnarray*}
D\equiv \ell^a\,\partial_a =\partial_r,
\end{eqnarray*}
\begin{eqnarray*}
\Delta\equiv n^a\,\partial_a ={r^2+a^2\over R^2}\,\partial_u-{H\over
2}\,\partial_r+{a\over
R^2}\partial_\phi,
\end{eqnarray*}
\begin{equation}
\delta\equiv m^a\,\partial_a ={1\over\surd 2\,R}\,\Big\{ia\,{\rm
sin}\,\theta\,\partial_u+\partial_\theta +{i\over{\rm
sin}\,\theta}\,\partial_\phi\Big\},
\end{equation}
\begin{eqnarray*}
\overline\delta\equiv \overline m^a\,\partial_a ={1\over\surd
2\,\overline R}\,\Big\{-ia\,{\rm
sin}\,\theta\,\partial_u+\partial_\theta -{i\over{\rm
sin}\,\theta}\,\partial_\phi\Big\}.
\end{eqnarray*}
By taking the exterior derivative of basis one-forms (3.1), we
calculate the spin coefficients from the Cartan's
equations of structure written in Newman-Penrose spin coefficients [19]:
\begin{eqnarray*}
\kappa=\sigma=\lambda=\epsilon=0,
\end{eqnarray*}
\begin{eqnarray*}
\rho=-{1\over\overline R},\;\;\;\;
\mu=-{H(u,r,\theta)\over{2\overline R}},
\end{eqnarray*}
\begin{eqnarray*}
\alpha={(2ai-R\,{\rm cos}\,\theta)\over{2\surd 2\overline
R\,\overline R\,{\rm sin}\,\theta}},\;\;\; \beta={{\rm
cot}\,\theta\over2\surd 2R},
\end{eqnarray*}
\begin{equation}
\pi={i\,a\,{\rm sin}\,\theta\over{\surd 2\overline R\,\overline
R}},\;\;\; \tau=-{i\,a\,{\rm sin}\,\theta\over{\surd 2R^2}},
\end{equation}
\begin{eqnarray*}
\gamma={1\over{\surd 2\,\overline R\,R^2}}\,
\Big [(r - M - r\,M_{,r} + e\,e_{,r})\, \overline R -
\Delta^*\Big ],
\end{eqnarray*}
\begin{eqnarray*}
\nu={1\over{\surd 2\,\overline R\,R^2}}\,\Big [i\,a\,{\rm
sin}\,\theta(r\,M_{,u}-e\,e_{,u})- (r\,M_{,\theta}-e\,e_{,\theta})
\Big ], \;\;
\end{eqnarray*}
where $\Delta^*=r^2-2rM(u,r,\theta)+a^2+e^2(u,r,\theta)$ and the
function $H(u,r,\theta)$ is given in (2.9). From these NP spin
coefficients we conclude that the transformed metric (2.7) with
the functions $M$ and $e$ of three variables $u,r,\theta$
possesses, in general, a geodesic $(\kappa=\epsilon=0)$, shear
free $(\sigma=0)$, expanding $(\hat{\theta} \neq 0)$ and rotating
$(\omega^{*2} \neq 0)$  null vector $\ell_a$ [20] where
\begin{equation}
\hat{\theta}\equiv -{1\over2}(\rho + \overline\rho) ={r\over R^2},
\end{equation}
\begin{equation}
\omega^{*2}\equiv-{1\over4}(\rho - \overline\rho)^2
=-{{a^2\cos^2\theta}\over {R^2\,R^2}}.
\end{equation}

Further we calculate the Weyl scalars:
\begin{eqnarray*}
\psi_0=\psi_1=0,
\end{eqnarray*}
\begin{eqnarray*}
\psi_2={1\over\overline R\overline R\,R^2}\left[(-RM+e^2)+\overline
R\,(rM_{,r}-ee_{,r})+{1\over 6}\,\overline R\,\overline
R\,\Big(-2M_{,r}-rM_{,rr}+e^2_{,r}+ee_{,rr}\Big)\right],
\end{eqnarray*}
\begin{eqnarray}
\psi_3 &=&{-1\over{2\surd 2\overline R\overline R\,R^2}}\,
\Big[4\Big\{i\,a\,{\rm sin}\,\theta\,(rM_{,u}-ee_{,u})
-(rM_{,\theta}-ee_{,\theta})\Big\}  \nonumber \\
&& + {\overline R}\,\Big\{i\,a\,{\rm
sin}\,\theta(r\,M_{,u}-e\,e_{,u})_{,r} -
(r\,M_{,\theta}-e\,e_{,\theta})_{,r}\Big\}\Big],
\end{eqnarray}
\begin{eqnarray*}
\psi_4 &=&{1\over{2\overline R\,\overline R\,R^2}}\,\Big [a^2{\rm
sin}^2\theta\,(rM_{,u}-ee_{,u})_{,u}+2ia\,{\rm
sin}\,\theta\,(rM_{,u}-ee_{,u})_{,\theta}-(rM_{,\theta}-
ee_{,\theta})_{,\theta}\Big ]    \nonumber  \\
&&- {{r\,a^2{\rm sin}^2\theta}\over{\overline R\,\overline
R\,R^2\,R^2}}\,(rM_{,u}-ee_{,u}) -{{2\,r\,a\,{\rm sin}^2\theta-
R^2{\rm cos}\,\theta}\over{2\overline R\,\overline R\,R^2\,R^2{\rm
sin}\,\theta}}\,(r\,M_{,\theta} -e\,e_{,\theta}). \;\;
\end{eqnarray*}

\vspace*{.15in} The non-vanishing of the Weyl scalars $(\psi_2
\neq \psi_3 \neq \psi_4 \neq 0)$ means that the metric (2.7) is an
algebraically special in the Petrov classification. In the
expression of $\psi_2$ it is found in general that there is no
differential terms of $M(u,r,\theta)$ and $e(u,r,\theta)$ with
respect to $u$ and $\theta$. This leads that for a static rotating
metric with the mass $M(r)$ and the charge $e(r)$, the spacetime
metric will be a Petrov type D $(\psi_2\neq 0, \psi_3=\psi_4=0)$,
whose one of the repeated principal null vectors  is a geodesic,
shear free, expanding  (3.4) and rotating (3.5) vector $\ell_a$.

The Ricci scalars of the metric (2.7) are obtained as follows:
\begin{eqnarray*}
\phi_{00}=\phi_{01}=\phi_{10}=\phi_{20}=\phi_{02}=0,
\end{eqnarray*}
\begin{eqnarray*}
\phi_{11}={1\over{4\,R^2\,R^2}}\,\Big[ 2\,e^2
+4r(r\,M_{,r}-ee_{,r})
+R^2\Big(-2M_{,r}-r\,M_{,r\,r}+e^2_{,r}+ee_{,r\,r}\Big)\Big],
\end{eqnarray*}
\begin{eqnarray*}
\phi_{12}&=&{1\over{2\surd 2\,R^2\,R^2}}\,\Big[i\,a\,{\rm
sin}\,\theta \Big\{(R\,M_{,u}-2\,e\,e_{,u})-
(r\,M_{,r}-e\,e_{,r})_{,u}\overline R\Big\} \nonumber \\
&&+\Big\{(R\,M_{,\theta}-2\,e\,e_{,\theta})-(
r\,M_{,r}-e\,e_{,r})_{,\theta}\,\overline R\Big\}\Big],
\end{eqnarray*}
\begin{eqnarray}
\phi_{21}&=&-{1\over{2\surd 2\,R^2\,R^2}}\,\Big [i\,a\,{\rm
sin}\,\theta\Big\{(\overline R\,M_{,u}-2\,e\,e_{,u})-
(r\,M_{,r}-e\,e_{,r})_{,u}\,R\Big\} \nonumber \\
&&+\Big\{(\overline R\,M_{,\theta}-2\,e\,e_{,\theta})
-(r\,M_{,r}-e\,e_{,r})_{,\theta}\,R\Big\}\Big],
\end{eqnarray}
\begin{eqnarray*}
\phi_{22}&=&-{1\over{2\,R^2\,R^2}}\,\Big [
2r(r\,M_{,u}-e\,e_{,u})-{\rm
cot}\,\theta(rM_{,\theta}-e\,e_{,\theta}) \nonumber \\
&& + a^2\,{\rm sin}^2\theta(r\,M_{,u}
    -e\,e_{,u})_{,u}-(rM_{,\theta}-e\,e_{,\theta})_{,\theta}\Big ],
\end{eqnarray*}
\begin{eqnarray*}
\Lambda={1\over{12\,R^2}}\,\Big(2M_{,r}+
r\,M_{,r\,r}-e^2_{,r}-ee_{,r\,r}\Big).\;\;
\end{eqnarray*}

\vspace*{.25in}

Here we observe that the general expressions of $\phi_{11}$ and
$\Lambda$ do not involve any differential terms with respect to
$u$ and $\theta$, although the functions $M$ and $e$ of the metric
(2.7) are functions of three variables $u,r,\theta$. The vanishing
of $\phi_{00}$ suggests the possibility that the transformed
metric (2.7) does not possess a perfect fluid
$T_{ab}=(\rho^*+p)u_au_b-p\,g_{ab}$ as
$\phi_{00}=2\phi_{11}=\phi_{22}=-K(\rho^*+p)/4$,
$\Lambda=K(3p-\rho^*)/24$ with a time-like vector
$u^a=(\ell^a+n^a)/\surd 2$ [21,22]. It is also worth mentioning
that for a static rotating metric with $M(r)$ and $e(r)$, the
Ricci scalars $\phi_{12}$ and $\phi_{22}$ will be vanished.

\vspace{.15in}

   Then the Einstein's tensor is computed from these Ricci
scalars (3.7) as follows
\begin{eqnarray}
G_{ab}&=&-2\,\phi_{22}\,\ell_a\,\ell_b-
4\,\phi_{11}\,\{\ell_{(a}\,n_{b)} +m_{(a}\overline
m_{b)}\}-6\,\Lambda\,g_{ab} \cr &&+
4\,\phi_{12}\,\ell_{(a}\,\overline m_{b)} +
4\,\phi_{21}\,\ell_{(a}\,m_{b)},
\end{eqnarray}
where $2\ell_{(a}\,n_{b)}$=$\ell_{a}\,n_{b} + n_{a}\,\ell_{b}$.
For non-rotating fields ($a=0$), the Ricci scalars $\phi_{12}$,
$\phi_{21}$ will vanish and this Einstein's tensor will reduce to
that presented by Glass and Krisch [23]. From the Einstein's
equations:
\begin{equation}
G_{ab}\equiv R_{ab} - {1\over 2}\,R\,g_{ab} = - K\,T_{ab}
\end{equation}
we obtain the null density $\mu^*$, the matter density $\rho^*$,
the pressure $p$ as well as the rotation function $\omega$ as
\begin{equation}
K\,\mu^* = 2\,\phi_{22},\;\;  K\,\omega = - 2\,\phi_{12},\;\;  \\
K\,\rho^* =  2\,\phi_{11} + 6\,\Lambda,\;\;  K\,p =
2\,\phi_{11} - 6\,\Lambda
\end{equation}
where the Ricci scalars $\phi_{11}$, $\phi_{12}$,
$\phi_{22}$, $\Lambda$ are given in (3.7).

To have the two-rotating fluid description we shall introduce a
time-like unit vector $u^a$ and three unit space-like vectors
$v^a$, $w^a$, $z^a$ such that
\begin{eqnarray*}
u_a={1\over\surd 2}(\ell_a+n_a),\\\
v_a={1\over\surd 2}(\ell_a-n_a),
\end{eqnarray*}
\begin{equation}
w_a={1\over\surd 2}(m_a+\overline m_a),\\\
z_a=-{i\over\surd 2}(m_a-\overline m_a)
\end{equation}
with the normalization conditions $u_a\,u^a = 1$, $v_a\,v^a$ =
$w_a\,w^a$ = $z_a\,z^a = -1$. Then the explicit forms of these
unit vectors are as follows
\begin{eqnarray*}
u_a={1\over\surd 2}\,\Big\{(1+{1\over 2}\,H)\,\delta^1_a+
\delta^2_a
-(1+{1\over 2}\,H)\,a\,{\rm sin}^2\theta\,\delta^4_a\Big\},
\end{eqnarray*}
\begin{eqnarray*}
v_a={1\over\surd 2}\,\Big\{(1-{1\over 2}\,H)\,\delta^1_a-
\delta^2_a
-(1-{1\over 2}\,H)\,a\,{\rm sin}^2\theta\,\delta^4_a\Big\},
\end{eqnarray*}
\begin{equation}
w_a=-{1\over R^2}\,\Big\{-a^2{\rm sin}\,\theta\,{\rm
cos}\,\theta\,\delta^1_a+r\,R^2\delta^3_a +a\,(r^2+a^2){\rm
sin}\,\theta\,{\rm cos}\,\theta\,\delta^4_a\Big\},
\end{equation}
\begin{eqnarray*}
z_a={1\over R^2}\,\Big\{a\,r\,{\rm
sin}\,\theta\,\delta^1_a+a\,{\rm cos}\,\theta\,R^2\delta^3_a
-r\,(r^2+a^2){\rm sin}\,\theta\,\delta^4_a\Big\},
\end{eqnarray*}
The metric tensor $g_{ab}$ can be expressed in these unit vectors
\begin{equation}
g_{ab}=u_a\,u_b-v_a\,v_b-w_a\,w_b-z_a\,z_b.
\end{equation}
Thus the Einstein's equations are written in two-fluid system
\begin{eqnarray}
G_{ab} &=& -K[\mu^*\,\ell_a\,\ell_b+
\rho^*\,(u_a\,u_b-v_a\,v_b)+p\,(w_a\,w_b+z_a\,z_b) \cr
& & + (\omega+\overline\omega)\,\{u_{(a}\,w_{b)}
+v_{(a}\,w_{b)}\}-i(\omega-\overline\omega)\,\{u_{(a}\,z_{b)}
+v_{(a}\,z_{b)}\}],
\end{eqnarray}
where $\mu^*$, $\rho^*$, $p$ and $\omega$ are related with the
Ricci scalars given in (3.7) as:
\begin{eqnarray*}
K\,\mu^* &=&-{1\over R^2\,R^2}\,\Big\{
2r(r\,M_{,u}-e\,e_{,u})-{\rm
cot}\theta(rM_{,\theta}-e\,e_{,\theta}) \nonumber \\
&& + a^2\,{\rm sin}^2\theta(r\,M_{,u}
-e\,e_{,u})_{,u}-(rM_{,\theta}-e\,e_{,\theta})_{,\theta}\Big\}
\end{eqnarray*}
\begin{eqnarray*}
K\,\rho^* =  {1\over
R^2\,R^2}\Big\{e^2+2\,r(r\,M_{,r}-e\,e_{,r})\Big\}
\end{eqnarray*}
\begin{equation}
K\,p =  {1\over R^2\,R^2}\Big\{e^2+2\,r(r\,M_{,r}-e\,e_{,r})
-R^2(2M_{,r}+r\,M_{,r\,r}-e^2_{,r}-e\,e_{,r\,r})\Big\},
\end{equation}
\begin{eqnarray*}
K\,\omega &=&-{1\over \surd 2\,R^2\,R^2}\,\Big [i\,a\,{\rm
sin}\,\theta \Big\{(R\,M_{,u}-2\,e\,e_{,u})-
(r\,M_{,r}-e\,e_{,r})_{,u}\overline R\Big\} \nonumber \\
&&+\Big\{(R\,M_{,\theta}-2\,e\,e_{,\theta})-(
r\,M_{,r}-e\,e_{,r})_{,\theta}\,\overline R\Big\}\Big ].
\end{eqnarray*}
The expression of null radiation density $\mu^*$ involves the
derivative of the functions $M(u,r,\theta)$ and $e(u,r,\theta)$
with respect to $u$ and $\theta$. Those of $\rho^*$ and $p$ are
with respect to $r$ only. However, the expression of the rotation
function $\omega$ is involved the derivative of mass and the
charge with respect to three variables $u$, $r$, $\theta$. From
the above equations, it is observed that the Einstein's tensor
(3.14) of the rotating fluid reduce to those of Glass and Krisch
[23] of {\sl non-rotating} string fluid when $a=e=0$ and
$M=M(u,r)$. In the non-rotating Vaidya-type radiation null fluid,
the null density $\mu^*$ takes the form $\mu^*=-2M_{,u}/Kr^2$.
This shows that $\mu^*$ is always negative, since $\partial
M/\partial u$ is positive [24].

\section{Stress-energy tensor and energy conditions}
\setcounter{equation}{0}
\renewcommand{\theequation}{4.\arabic{equation}}

From the Einstein tensor (3.8) and the relations (3.10) of Ricci
scalars with $\mu^*$, $\rho^*$, $p$, $\omega$ we can introduce the
total energy momentum tensor (EMT) for a rotating fluid as
follows:
\begin{eqnarray}
T_{ab} &=& T^{(\rm n)}_{ab} +T^{(\rm m)}_{ab} \cr
&=&\mu^*\,\ell_a\,\ell_b+
2\,\rho^*\,\ell_{(a}\,n_{b)}
+2\,p\,m_{(a}\overline m_{b)} \cr
& & + 2\,\omega\,\ell_{(a}\,\overline m_{b)} +
2\,\overline\omega\,\ell_{(a}\,m_{b)}
\end{eqnarray}
where the EMTs for the rotating null fluid as well as that of the
rotating matter are given respectively below:
\begin{equation}
T^{(\rm n)}_{ab}= \mu^*\,\ell_a\,\ell_b +
\omega\,\ell_{(a}\,\overline
m_{b)}+\overline\omega\,\ell_{(a}\,m_{b)}
\end{equation}
\begin{equation}
T^{(\rm m)}_{ab}=2\,(\rho^*+p)\,\ell_{(a}\,n_{b)}
- p\,g_{ab} + \omega\,\ell_{(a}\,\overline m_{b)} +
\overline\omega\,\ell_{(a}\,m_{b)},
\end{equation}
where $\overline\omega$ is the complex conjugate of $\omega$. When
$\omega = 0$ initially, these EMTs is similar to those introduced
by Husain [25] in the case of non-rotating fluid.

In General Relativity the stress-energy tensor represents the
matter that describes the gravitation in the space-time geometry
through Einstein's field equations. From the conditions of Ricci
scalar $\phi_{00}=0$, obtained above for the space-time geometry
(2.7), it may be concluded that the stress-energy tensor given
above does not, in general describe a perfect fluid, as for a
perfect fluid the Ricci scalar $\phi_{00}=-K(\rho^*+p)/4$ must not
vanish. Hence, it will be interesting to study the nature of the
energy conditions for rotating non-perfect fluid given in (4.1).
When the rotation factor $\omega$ vanishes in (4.2), this fluid
may be thought of null radiation fluid of non-rotating Vaidya
space-time. So we refer to this rotating ($\omega\neq 0$) null
radiation fluid as rotating Vaidya type radiating fluid, shortly
rotating Vaidya fluid (4.2).

As the $T_{ab}$ does not include the perfect fluid, it seems that
the stress-energy tensor can represent the interaction of rotating
Vaidya fluid with rotating non-perfect fluid. Since there is a
coupling term of the rotation scalar $a$ with $\partial M/\partial
u$ in the expression of $\omega$ appearing in $T_{ab}$, the energy
condition of this $T_{ab}$ satisfying Einstein's field equations
will be a new area to discuss in the classical General Relativity.
For this purpose, we write the matter part $T^{(\rm m)}_{ab}$ of
(4.1) in terms of time-like $u^a$ as well as space-like vectors
$v^a$, $w^a$, $z^a$ given in (3.12) as
\begin{eqnarray}
T_{ab} &=&\mu^*\,\ell_a\,\ell_b+
(\rho^*+p)\,(u_a\,u_b-v_a\,v_b)-p\,g_{ab} \cr
& & + (\omega+\overline\omega)\,\{u_{(a}\,w_{b)}
+v_{(a}\,w_{b)}\}-i(\omega-\overline\omega)\,\{u_{(a}\,z_{b)}
+v_{(a}\,z_{b)}\},
\end{eqnarray}
and its trace is $T\equiv T_{ab}g^{ab}=2(\rho^*-p)$, which is
different from that of a perfect fluid. This trace will be
vanished when $\rho*=p$. This means that the matter part $T^{(\rm
m)}_{ab}$ of the stress-energy tensor may be that of
electromagnetic field whose trace is zero. [The stress-energy
tensor for a non-rotating perfect fluid is $T^{(\rm {pf})}_{ab} =
(\rho^*+p)u_a\,u_b-p\,g_{ab}$ with unit time-like vector $u_a$ and
trace $T^{(\rm {pf})}=\rho^*-3p$, which will be zero when
$\rho*=3p$].

To study the energy conditions of the energy-momentum tensor, we
consider a time-like observer with its four-velocity vector
$U_{a}$ [26,27]
\begin{equation}
U_{a}=\hat{\alpha}u_a+
\hat{\beta}v_a+\hat{\gamma}w_a+\hat{\delta}z_a,
\end{equation}
where $\hat{\alpha}$, $\hat{\beta}$, $\hat{\gamma}$ and
$\hat{\delta}$ are arbitrary constants. The four-velocity vector
$U_a$ is subjected to the condition that
\begin{equation}
U^aU_a =
\hat{\alpha}^2-\hat{\beta}^2-\hat{\gamma}^2-\hat{\delta}^2 \geq 0.
\end{equation}
Now, $T_{ab}U^aU^b$ will represent the energy density as measured by the
time-like observer with the unit tengent vector $U^a$.
Then the energy conditions [36] are the following:

(a) {\it Weak energy condition}: The energy momentum tensor obeys
the inequality $T_{ab}U^aU^b\geq0$ for any timelike vector $U^a$
i,e., $T_{ab}U^aU^b\geq0$ implies that
\begin{eqnarray*}
(i)\:\: {\mu^*\over 2} + \rho^* \geq 0,\:\:\:\:
(ii)\:\: \pm{\mu^*\over 2} + \rho^* +p \geq 0,\:\:\:\: (iii)\:\: {3\mu^*\over 2}+
\rho^* + p\geq 0,
\end{eqnarray*}
\begin{equation}
(iv)\,\, \big({\mu^*\over 2}+\rho^* + p\big) \pm
(\omega+\overline\omega)\geq 0,\:\: (v)\,\, \big({\mu^*\over 2}+\rho^*+
p\big) \pm i\,(\omega-\overline\omega) \geq 0.
\end{equation}

(b) {\it Strong energy condition}: The Ricci tensor for $T_{ab}$
(4.4) satisfies the inequality $R_{ab}U^aU^b\geq 0$ for
all time-like vector $U^a$, i.e. $T_{ab}U^aU^b\geq {1\over2}T$, which implies
that
\begin{eqnarray*}
(i)\:\: {\mu^*\over 2} + p \geq 0,\:\:\:\:
(ii)\:\: \pm{\mu^*\over 2} + \rho^* +p \geq 0,\:\:\:\: (iii)\:\: {3\mu^*\over 2}+
\rho^* + p\geq 0,
\end{eqnarray*}
\begin{equation}
(iv)\,\, \big({\mu^*\over 2}+\rho^* + p\big) \pm
(\omega+\overline\omega)\geq 0,\:\: (v)\,\, \big({\mu^*\over 2}+\rho^*+
p\big) \pm i\,(\omega-\overline\omega) \geq 0.
\end{equation}

(c) {\it Dominant energy condition}: For all future directed,
time-like vector $U^a$, $T_{ab}U^b$ should be a future directed
non-space like vector. This energy condition is equivalent to
\begin{eqnarray*}
(i) \,\, \mu^*\rho^*+\rho^{*2}-(f^2+g^2)\geq 0,
\end{eqnarray*}
\begin{eqnarray*}
(ii) \,\, -\mu^*\rho^*+\rho^{*2}-p^2+(f^2+g^2)\geq 0,\,\,
(iii)\,\, 3\mu^*\rho^*+\rho^{*2}-p^2-3(f^2+g^2)\geq 0,
\end{eqnarray*}
\begin{equation}
(iv)\,\, \{\mu^*\rho^*+\rho^{*2}- p^2-(f^2+g^2)\}\pm
2f(\rho^*-p)\geq 0,
\end{equation}
\begin{eqnarray*}
(v)\,\, \{\mu^*\rho^*+\rho^{*2}- p^2-(f^2+g^2)\}\pm
2g(\rho^*-p)\geq 0,
\end{eqnarray*}
where $2f=\omega+\overline\omega$ and $2\,i\,g =
\omega-\overline\omega$. These are the energy conditions satisfied
by the energy-momentum tensor (4.4) in general. We find that the
rotation function $\omega$ is involved in all the three energy
conditions.

\section{Rotating solutions recovered from the general solutions}
\setcounter{equation}{0}
\renewcommand{\theequation}{5.\arabic{equation}}

In the above section we present the full expressions of NP spin
coefficients (3.3) the Weyl scalars (3.6) and the Ricci scalars
(3.7) with arbitrary functions $M$ and $e$ of three coordinate
variables $u,r,\theta$. These NP quantities are so transparent
that these quantities can explain the nature of solutions of
Einstein's field equations discussed in this paper. For example,
the NP spin coefficients (3.3) easily explain that the metric
(2.7) admits a null vector $\ell^a$ which is geodesic, shear free,
expanding (3.4) as well as rotating (3.5), In this section, with
the help of NP quantities given in (3.3), (3.6), (3.7), we shall
give known examples of rotating solutions like Kerr-Newman,
rotating Vaidya [4] and rotating Vaidya-Bonnor [10]. In the next
section we shall combine these solutions with other rotating
solutions in order to derive new embedded rotating solutions.

\subsection{\it Kerr-Newman solution: $e$ = $M$ =
constant, $a\neq 0$}

When $e$ = $M$ = constant, $a\neq 0$, the equation (3.10)
reduces to the Kerr-Newman solution
\begin{eqnarray*}
\mu^* = \omega = 0
\end{eqnarray*}
\begin{equation}
\rho^* = p = {e^2\over K\,R^2R^2},
\end{equation}
and the only existing Weyl scalar is
\begin{eqnarray*}
\psi_2={1\over\overline R\,\overline R\,R^2}(e^2-R\,M).
\end{eqnarray*}
Then the total energy momentum tensor takes the form
\begin{eqnarray}
T_{ab} &=& \rho^*\,(\ell_{a}\,n_{b}+n_a\,\ell_b)
+p\,(m_{a}\overline m_{b} + \overline m_a\,m_b) \cr
&=& (e^2/KR^2R^2)\{(\ell_{a}\,n_{b}+n_a\,\ell_b)
+(m_{a}\overline m_{b} + \overline m_a\,m_b)\},
\end{eqnarray}
which is the EMT for non-null electromagnetic field with Maxwell
scalar
\begin{equation}
\phi_1 \equiv {1\over 2}F_{ab}(\ell^a\,n^b + \overline
m^a\,m^b)={e\over \surd (2K)\overline R\,\overline R}
\end{equation}
for Kerr-Newman solution. Here is the birth place of Kerr-Newman
solution, originally applied the Newman-Janis algorithm by Newman
{\sl et. al.} [12] to generate this well known rotating solution
from the non-rotating Reissner-Nordstrom `seed' solution. The line
element is
\begin{eqnarray}
ds^2&=&\{1-(2rM-e^2)R^{-2}\}\,du^2+2du\,dr \cr
&&+2aR^{-2}(2rM-e^2){\rm sin}^2\theta\,du\,d\phi
-2a\,{\rm sin}^2\theta\,dr\,d\phi \cr
&&-R^2d\theta^2-\{(r^2+a^2)^2- \Delta^*a^2\,{\rm
sin}^2\theta\,\}R^{-2}{\rm
sin}^2\theta\,d\phi^2,
\end{eqnarray}
where $\Delta^*=r^2-2rM+a^2+e^2$.
The charged Kerr-Newman black hole has an {\sl
external event horizon} at $r_{+}=M+\surd
{(M^2-a^2-e^2)}$\, and an {\sl internal Cauchy horizon} at
$r_{-}=M-\surd {(M^2-a^2-e^2)}$. The {\sl stationary limit}
surface $g_{uu}>0$ of the rotating black hole i.\,e.
$r=r_e(\theta)=M+\surd {(M^2-a^2{\rm cos}^2\theta-e^2)}$
does not coincide with the event horizon at $r_{+}$
thereby producing the {\sl ergosphere}. This
stationary limit coincides with the event horizon at the
poles $\theta=0$ and $\theta=\pi$ [20].
Naturally, this solution
includes Kerr ($e=0$), Reissner-Nordstrom ($a=0$, $e\neq0$) as
well as Schwarzschild ($a=e=0$) solutions.

\subsection{\it Rotating Vaidya solution:
$M = M(u)$, $a\neq 0$. $e=0$}

In this case the energy momentum tensor (4.2) takes
\begin{equation}
T_{ab}= \mu^*\,\ell_a\,\ell_b +
\omega\,\ell_{(a}\,\overline
m_{b)}+\overline\omega\,\ell_{(a}\,m_{b)}
\end{equation}
where the null density $\mu^*$ and the rotation function
$\omega$ in (3.15) become
\begin{eqnarray*}
K\,\mu^* = -{1\over R^2\,R^2}\Big\{2r^2M_{,u}+a^2r{\rm
sin}^2\theta\,M_{,uu}\Big\},
\end{eqnarray*}
\begin{equation}
K\,\omega =-{1 \over{\surd
2\,\overline R\,R^2}}\,i\,a\,{\rm sin}\,\theta\,M_{,u},
\end{equation}
and the Weyl scalars are
\begin{eqnarray*}
\psi_2=-{M\over\overline R\,\overline
R\,\overline R},
\end{eqnarray*}
\begin{equation}
\psi_3 =-{i\,a\,{\rm sin}\theta\over 2\surd 2\overline
R\,\overline R\,R^2}
\Big\{4\,rM_{,u}+\overline R\,M_{,u}\Big\},
\end{equation}
\begin{equation}
\psi_4 ={{a^2\,r\,{\rm sin}^2\theta}\over 2\overline
R\,\overline R\,R^2\,R^2}\,\Big\{R^2\,M_{,uu}
-2r\,M_{,u}\Big\}.
\end{equation}
From the above we observe that $\omega$, $\psi_3$, $\psi_4$ will
vanish when $a=0$, and the EMT will be that of the original
non-rotational radiating Vaidya metric [28] with $\mu^*
=-{2\,M_{,u}/K\,r^2}$. The line element of this rotating metric is
\begin{eqnarray}
ds^2&=&\{1-2rM(u)R^{-2}\}\,du^2+2du\,dr\cr
&&+4arM(u){\rm sin}^2\theta\,R^{-2}\,du\,d\phi
-2a\,{\rm sin}^2\theta\,dr\,d\phi \cr
&&-R^2d\theta^2-\{(r^2+a^2)^2-\Delta^*\,a^2\,{\rm
sin}^2\theta\}R^{-2}{\rm
sin}^2\theta\,d\phi^2,
\end{eqnarray}
where $\Delta^*=r^2-2rM(u)+a^2$. This metric represents a
non-stationary rotating solution of Einstein's equations
possessing an energy-momentum tensor (5.5) for a rotating null
radiating fluid, and can describe a non-stationary {\it rotating}
black hole if $M(u)>a$. The involvement of $\omega$ in the
energy-momentum tensor (5.5) indicates that the null fluid is a
rotating Vaidya null fluid. When $M(u)$ = constant initially, this
metric would reduce to rotating vacuum Kerr solution with
vanishing $\mu^*$ and $\omega$ in (5.6).

Carmeli and Kaye [4] studied the metric (5.9) after considering
the mass $M$ of the Kerr solution as a function of coordinate $u$.
That is why, they referred to the metric (5.9) as the
variable-mass Kerr solution ( see also in [29,30]) and discussed
the properties of the metric using the NP quantities. Carmeli [29]
referred to these $\omega$ terms as residues of the black hole.
However, we refer to the metric (5.9) as rotating Vaidya solution.
In the next section this metric will be combining smoothly with
the usual Kerr-Newman solution as rotating Kerr-Newman-Vaidya
black hole.

\subsection{\it Rotating Vaidya-Bonnor solution:
$M=M(u)$, $a\neq 0$. $e=e(u)$}

In this case the energy momentum tensor takes
\begin{eqnarray}
T_{ab}= \mu^*\,\ell_a\,\ell_b +
2\,\rho^*\,\{\ell_{(a}\,n_{b)}
+m_{(a}\overline m_{b)}\}
+ 2\,\omega\,\ell_{(a}\,\overline m_{b)} +
2\,\overline\omega\,\ell_{(a}\,m_{b)}
\end{eqnarray}
where
\begin{eqnarray*}
\mu^* =-{1\over
K\,R^2\,R^2}\Big\{2\,r\,(r\,M_{,u}-e\,e_{,u})+a^2{\rm
sin}^2\theta\,(r\,M_{,u}-e\,e_{,u})_{,u}\Big\},
\end{eqnarray*}
\begin{equation}
\rho^* = p = {e^2(u)\over K\,R^2\,R^2},
\end{equation}
\begin{eqnarray*}
\omega ={-i\,a\,{\rm sin}\,\theta\,\over{\surd
2\,K\,R^2\,R^2}}\,\Big\{R\,M_{,u}-2e\,e_{,u}\Big\},
\end{eqnarray*}
and the Weyl scalars are
\begin{eqnarray*}
\psi_2={1\over\overline R\,\overline R\,R^2}\Big(e^2 -R\,M\Big)
\end{eqnarray*}
\begin{equation}
\psi_3 ={-i\,a\,{\rm sin}\theta\over 2\surd 2\overline
R\,\overline R\,R^2}
\Big\{4\,(rM_{,u}-e\,e_{,u})+\overline R\,M_{,u}\Big\},
\end{equation}
\begin{eqnarray*}
\psi_4 ={{a^2\,{\rm sin}^2\theta}\over 2\overline
R\,\overline R\,R^2\,R^2}\,\Big\{
R^2\,(r\,M_{,u}-e\,e_{,u})_{,u}-2r\,(r\,M_{,u}-e\,e_{,u})\Big\}.
\end{eqnarray*}
The line element will be in the form
\begin{eqnarray}
ds^2&=&[1-\{2rM(u)-e^2(u)\}R^{-2}]\,du^2+2du\,dr \cr
&&+2aR^{-2}\{2rM(u)-e^2(u)\}{\rm sin}^2\theta\,du\,d\phi
-2a\,{\rm sin}^2\theta\,dr\,d\phi \cr
&&-R^2d\theta^2-\{(r^2+a^2)^2- \Delta^*a^2\,{\rm
sin}^2\theta\,\}R^{-2}{\rm
sin}^2\theta\,d\phi^2,
\end{eqnarray}
where $\Delta^*=r^2-2rM(u)+a^2+e^2(u)$. This solution will
describe a black hole when $M(u)>a^2+e^2(u)$ and has
$r_{\pm}=M(u)^*\pm \surd {\{M^2(u)-a^2-e^2(u)\}}$ as the roots of
the equation $\Delta^*=0$. So the rotating Vaidya-Bonnor solution
has an {\it external event horizon} at $r_{+}= M(u)+\surd
{\{M^2(u)-a^2-e^2(u)\}}$ and an {\it internal Cauchy horizon} at
$r_{-}= M(u)-\surd {\{M^2(u)-a^2-e^2(u)\}}$. The non-stationary
limit surface $g_{uu}>0$ of the rotating black hole {\it i.e.}
$r\equiv r_e(u,\theta)=M(u)+\surd {\{M^2(u)-a^2{\rm
cos}^2\theta-e^2(u)\}}$ does not coincide with the event horizon
at $r_+$, thereby producing the {\it ergosphere}. The rotating
Vaidya-Bonnor metric (5.13) can be written in Kerr-Schild form on
the rotating Vaidya null radiating background as
\begin{equation}
g_{ab}^{\rm VB}=g_{ab}^{\rm V} +2Q(u,r,\theta)\ell_a\ell_b
\end{equation}
where
\begin{equation}
Q(u,r,\theta) = {e^2(u)\over2\,R^2}.
\end{equation}
Here, $g_{ab}^{\rm V}$ is the rotating Vaidya metric (5.9) and
$\ell_a$ is geodesic, shear free, expanding and rotating null
vector for both $g_{ab}^{\rm V}$ as well as $g_{ab}^{\rm VB}$ and
given in (2.8). The Kerr-Schild form (5.14) may be interpreted as
the existence of the electromagnetic field on the rotating Vaidya
null radiating background. If we set $M(u)$ and $e(u)$ are both
constant, this Kerr-Schild form may be that of Kerr-Newman black
hole. That is, the Kerr-Newman solution itself has the Kerr-Schild
form on the Kerr background with the same null vector $\ell_a$
(2.8).

From this rotating Vaidya-Bonnor metric,  we can clearly recover
the following solutions: (i) rotating Vaidya metric (5.9) when
$e(u)=0$, (ii) rotating charged Vaidya solution when $e(u)$
becomes constant, (iii) the Kerr-Newman solution (5.4) when $M(u)
= e(u)$ = constant and (iv) well-known non-rotating Vaidya-Bonnor
metric [31] when $a=0$. It is also noted that when $e=a=0$, the
null density of Vaidya radiating fluid takes the form $\mu^*
=-{2\,M_{,u}/K\,r^2}$. The non-rotating Vaidya null radiating
metric is of type $D$ in the Petrov classification of spacetime
whose one of the repeated principal null vectors, $\ell_a$ is a
geodesic, shear free, non-rotating with non-zero expansion [29],
while the rotating one is of algebraically special with a null
vector $\ell_a$ which is geodesic, shear free, rotating as well as
expanding. It is also noted that when $e=a=0$, the energy-momentum
tensor becomes that of the original non-rotational null-radiating
Vaidya fluid with $\mu^* =-{2\,M_{,u}/K\,r^2}$. It is noted that
the metric (5.13) can be seen in [10]. From (5.4) and (5.13) it is
observed that the rotating Vaidya-Bonnor solution is the
non-stationary version of Kerr-Newman black holes. That is, the
parameters $M$ and $e$ of Kerr-Newman solution are functions of
retarded time coordinate $u$ in rotating Vaidya-Bonnor
metric.

\section{\it Rotating solutions with $M=M(u,r)$,
$e(u,r,\theta)=0$} \setcounter{equation}{0}
\renewcommand{\theequation}{6.\arabic{equation}}

In this section we shall discuss the rotating solutions by
considering the function $M(u,r)$ of two variables $u,r$ only and
$e(u,r,\theta) = 0$, and derive new embedded rotating solutions.
We shall express all the rotating embedded solutions in
Kerr-Schild types of metrics in order to show them as solutions of
Einstein's field equations. In this case the energy momentum
tensor takes the form
\begin{eqnarray}
T_{ab}&=& \mu^*\,\ell_a\,\ell_b +
2\,\rho^*\,\ell_{(a}\,n_{b)}
+2\,p\,m_{(a}\overline m_{b)}+
2\,\omega\,\ell_{(a}\,\overline
m_{b)}+2\,\overline\omega\,\ell_{(a}\,m_{b)}
\end{eqnarray}
or in terms of unit vectors
\begin{eqnarray}
T_{ab}&=& \mu^*\,\ell_a\,\ell_b+
\rho^*\,(u_a\,u_b-v_a\,v_b)+p\,(w_a\,w_b+z_a\,z_b) \cr
& & + (\omega+\overline\omega)\,\{u_{(a}\,w_{b)}
+v_{(a}\,w_{b)}\}-i(\omega-\overline\omega)\,\{u_{(a}\,z_{b)}
+v_{(a}\,z_{b)}\},
\end{eqnarray}
where
\begin{eqnarray}
\mu^* &=&-{1\over K\,R^2\,R^2}\Big\{2r^2M_{,u} + a^2r\,{\rm
sin}^2\theta\,M_{,uu}\Big\},                             \cr
\rho^* &=& {2\,r^2\over K\,R^2\,R^2}\,M_{,r},                 \cr
p &=& -{1\over K}\,\Big\{{2\,a^2\,{\rm cos}^2\theta \over
R^2\,R^2}\,M_{,r}+{r\over R^2}\,M_{,r\,r}\Big\},              \cr
\omega &=&-{i\,a\,{\rm sin}\,\theta\over\surd
2\,K\,R^2\,R^2}\,\Big(R\,M_{,u}-r\,\overline R\,M_{,ur}\Big).
\end{eqnarray}
The line element will take the form
\begin{eqnarray}
ds^2&=&\Big\{1-2rM(u,r)R^{-2}\Big\}\,du^2+2du\,dr \cr
&&+4arM(u,r)R^{-2}\,{\rm sin}^2 \theta\,du\,d\phi -2a\,{\rm
sin}^2\theta\,dr\,d\phi \cr &&-R^2d\theta^2-\Big\{(r^2+a^2)^2
-\Delta^*a^2\,{\rm sin}^2\theta\Big\}\,R^{-2}{\rm
sin}^2\theta\,d\phi^2,
\end{eqnarray}
where $R^2=r^2+a^2{\rm cos}^2\theta$,
$\Delta^*=r^2-2rM(u,r)+a^2$
and the Weyl scalars are
\begin{eqnarray}
\psi_2&=&{1\over\overline R\,\overline
R\,R^2}\Big\{-R\,M+{\overline R\over 6}\,M_{,r}(4\,r+2\,i\,a\,{\rm
cos}\theta)-{r\over 6}\,\overline R\,\overline R\,M_{,rr}\Big\},
\cr \psi_3&=&-{i\,a\,{\rm sin}\theta\over{2\surd 2\overline
R\,\overline R\,R^2}}\, \Big\{(4\,r+\overline R\,)\,M_{,u}
 + r\,{\overline R}\,M_{,ur}\Big\}, \cr
\psi_4 &=&{{a^2\,r\,{\rm sin}^2\theta}\over 2\overline
R\,\overline R\,R^2\,R^2}\,\Big\{R^2\,M_{,uu}-2\,r\,M_{,u}\Big\}.
\end{eqnarray}
One can consider this rotating metric (6.4) along with the
stress-energy momentum tensor (6.1) or (6.2) and the Weyl scalars
as the extension of the non-rotating solutions discussed by Glass
and Krisch [23] and Husain [25].

\subsection{\it Rotating Husain's solution:
$M=M(u,r)$, $a\neq 0$}

Husain [25] has imposed one condition in the equation of state of
non-rotating null fluid that $p=k\rho^{*b}$ and obtain the
solution of the equation of state with $k\geq 1/2$ $b=1$. However,
due the present of the rotating factor $a$ in equation (6.3), one
cannot be able to get the solution. So we put $k=1$ and $b=1$.
Then, the equation to be solved takes a simple form
\begin{equation}
{M_{,r}\over r} = - {M_{,rr}\over 2},
\end{equation}
which gives the function $M(u,r)$
\begin{equation}
M(u,r) = f(u)-{1\over r}\,g(u).
\end{equation}
It can be treated as rotating Husain's solution for $p=k\rho^*$
with $k=1$. This rotating Husain's solution may be degenerated to
the rotating Vaidya-Bonnor solution presented above if one puts
$g(u)=e^2(u)/2$ in (6.7).

\subsection{\it Rotating Wang-Wu solutions}

Wang and Wu [32] have expanded the function $M(u,r)$ of (6.3) of
the non-rotating space in the power of $r$
\begin{equation}
M(u,r)= \sum_{n=-\infty}^{+\infty} q_n(u)\,r^n,
\end{equation}
where $q_n(u)$ are arbitrary functions of $u$. They consider the
above sum as an integral when the `spectrum' index $n$ is
continuous. In fact Wang and Wu technique is based on a linear
superposition that a linear superposition of mass function of
particular solutions is also a solution of Einstein's field
equations of non-rotating spacetime. Using the expression (6.8) in
equations (6.3) we can generate rotating solutions with Wang-Wu
functions as
\begin{eqnarray}
\mu^* &=&-{r\over K\,R^2\,R^2}\,\sum_{n=-\infty}^{+\infty}
\Big\{2\,q_n(u)_{,u}\,r^{n+1}+a^2{\rm
sin}^2\theta\,q_n(u)_{,uu}\,r^n\Big\},                    \cr
\rho^* &=& {2\,r^2\over K\,R^2\,R^2}\,\sum_{n=-\infty}^{+\infty}
n\,q_n(u)\,r^{n-1},                                           \cr
p &=& -{1\over K\,R^2}\,\sum_{n=-\infty}^{+\infty}
n\,q_n(u)\,r^{n-1}\Big\{{2\,a^2\,{\rm cos}^2\theta \over
R^2}+(n-1)\Big\},                                             \cr
\omega &=&-{i\,a\,{\rm sin}\,\theta\over\surd
2\,K\,R^2\,R^2}\,\sum_{n=-\infty}^{+\infty}\Big(R-n\overline
R\Big)\,q_n(u)_{,u}\,r^n.
\end{eqnarray}
Here one can observe that these rotating solutions with functions
$q_n(u)$ include many known as well as unknown rotating solutions
of Einstein's field equations in spherical symmetry as shown by
Wang and Wu in non-rotating cases [32]. The functions $q_n(u)$ in
(6.8) play a great role in generating new solutions whether
rotating or non-rotating. Therefore, we will hereafter refer to
these as Wang-Wu functions. Thus, rotating solutions can be
derived from these solutions as follows.

\subsubsection{\it Rotating monopole solution}

If one chooses the
functions $q_n(u)$ such that
\begin{eqnarray}
\begin{array}{cc}
q_n(u)=&\left\{\begin{array}{ll}
(b/2),&{\rm when }\;\;n=1\\
0, &{\rm when }\;\;n\neq 1
\end{array}\right.
\end{array}
\end{eqnarray}
where $b$ is constant, then from (6.8) one can obtain
\begin{eqnarray*}
M(u,r)=b\,r/2,\;\; \mu*=\omega=0,
\end{eqnarray*}
\begin{eqnarray*}
\rho^* =  {r^2\,b\over K\,R^2\,R^2},\;\;\;
p = -{b\,a^2{\rm cos}^2\theta\over K\,R^2\,R^2},
\end{eqnarray*}
with the energy momentum tensor
\begin{eqnarray*}
T_{ab}=2\,\rho^*\,\ell_{(a}\,n_{b)}
+2\,p\,m_{(a}\overline m_{b)}.
\end{eqnarray*}
The Weyl scalar takes the form
\begin{eqnarray*}
\psi_2=-{b\,r\over2\overline R\,\overline R\,\overline R}.
\end{eqnarray*}
The rotating monopole line element will be of the following form
\begin{eqnarray}
d
s^2&=&\Big\{1-b\,r^2R^{-2}\Big\}\,du^2+2du\,dr \cr
&&+2a\,b\,r^2R^{-2}\,{\rm sin}^2
\theta\,du\,d\phi -2a\,{\rm sin}^2\theta\,dr\,d\phi \cr
&&-R^2d\theta^2-\Big\{(r^2+a^2)^2
-\Delta^*a^2\,{\rm sin}^2\theta\Big\}\,R^{-2}{\rm
sin}^2\theta\,d\phi^2,
\end{eqnarray}
where $R^2=r^2+a^2{\rm cos}^2\theta$, and
$\Delta^*=r^2-b\,r^2+a^2$.

Because of the non-vanishing Weyl scalar $\psi_2$,  the rotating
monopole solution is stationary Petrov type $D$ whose one of the
repeated principal directions is geodesic, shear free, expanding
and rotating null vector $\ell_a$. The rotating monopole solution
has the non-zero pressure $p$, which leads the difference between
the rotating and the non-rotating monopole solutions. That is,
when $a=0$, one can obtain the non-rotating metric [32] with
$p=0$. To study this rotating monopole solution (6.11) will be of
interest. For example, one can easily embed Kerr-Newman black hole
in this rotating monopole space to study a different physical
nature of the black holes.

\subsubsection{\it Kerr-Newman solution}

We can choose the Wang-Wu functions $q_n(u)$ such that
\begin{eqnarray}
\begin{array}{cc}
q_n(u)=&\left\{\begin{array}{ll}
m,&{\rm when }\;\;n=0\\
-e^2/2, &{\rm when }\;\;n=-1\\
0, &{\rm when }\;\;n\neq 0, -1
\end{array}\right.
\end{array}
\end{eqnarray}
where $m$ and $e$ are constants. Then we obtain the function from
(6.8)
\begin{eqnarray*}
M(u,r)=m-e^2/2r,\;\;\: \mu*=\omega=0,
\end{eqnarray*}
\begin{equation}
\rho^* = p = {e^2\over K\,R^2\,R^2},
\end{equation}
and the Weyl scalar is
\begin{eqnarray*}
\psi_2={1\over\overline R\,\overline R\,R^2}(e^2-m\,R),
\end{eqnarray*}
which are the same as given (5.1).

\subsubsection{\it Rotating Vaidya-Bonnor solution}

The rotating Vaidya-Bonnor solution presented above, can also be
obtained from these rotating Wang-Wu solutions if we choose the
functions as
\begin{eqnarray}
\begin{array}{cc}
q_n(u)=&\left\{\begin{array}{ll}
f(u),&{\rm when }\;\;n=0\\
-h(u)^2/2, &{\rm when }\;\;n=-1\\
0, &{\rm when }\;\;n\neq 0, -1.
\end{array}\right.
\end{array}
\end{eqnarray}
Then the corresponding quantities are
\begin{eqnarray*}
M(u,r)=f(u)-h^2(u)/2r
\end{eqnarray*}
\begin{equation}
\rho^* = p = {h^2(u)\over K\,R^2\,R^2},
\end{equation}
\begin{eqnarray*}
\mu^* =-{1\over
K\,R^2\,R^2}\Big\{2\,r\,(r\,f(u)_{,u}-h\,h_{,u})+a^2{\rm
sin}^2\theta\,(r\,f(u)_{,u}-h\,h_{,u})_{,u}\Big\},
\end{eqnarray*}
\begin{equation}
\omega ={-i\,a\,{\rm sin}\,\theta\,\over{\surd
2\,K\,R^2\,R^2}}\,\Big\{R\,f(u)_{,u}-2h\,h_{,u}\Big\}.
\end{equation}
This solution includes the rotating Vaidya solution ($h(u) = 0$)
obtained above in (5.9). These subsections (6.2.2) and (6.2.3),
which are the repetition of the section (5.1) and (5.3) above,
show that the general rotating solutions with Wang-Wu functions
(6.8) can employ to derive the rotating Kerr-Newman as well as
rotating Vaidya-Bonnor solutions in a much simpler way.

\subsubsection{\it Kerr-Newman-Vaidya solution}

Wang and Wu [32] could combine the three non-rotating solutions,
namely monopole, de-Sitter and charged Vaidya solution to obtain a
new solution representing the non-rotating monopole-de
Sitter-Vaidya charged solutions. In the same way, we shall combine
the Kerr-Newman solution with the rotating Vaidya solution
obtained above in (5.9), if the Wang-Wu functions $q_n(u)$ are
chosen such that
\begin{eqnarray}
\begin{array}{cc}
q_n(u)=&\left\{\begin{array}{ll}
m+f(u),&{\rm when }\;\;n=0\\
-e^2/2, &{\rm when }\;\;n=-1\\
0, &{\rm when }\;\;n\neq 0, -1,
\end{array}\right.
\end{array}
\end{eqnarray}
where $m$ and $e$ are constants and $f(u)$ is the mass function of
rotating Vaidya solution (5.9). Thus, we obtain the function from
(6.8)
\begin{eqnarray*}
M(u,r)=m+f(u)-e^2/2r
\end{eqnarray*}
and using this in (6.9) we obtain other quantities
\begin{equation}
\rho^* = p = {e^2\over K\,R^2\,R^2},
\end{equation}
\begin{eqnarray*}
\mu^* =-{r\over
K\,R^2\,R^2}\Big\{2\,r\,f(u)_{,u}+a^2{\rm
sin}^2\theta\,f(u)_{,uu}\Big\},
\end{eqnarray*}
\begin{equation}
\omega ={-i\,a\,{\rm sin}\,\theta\,\over{\surd
2\,K\,\overline R\,R^2}}\,f(u)_{,u}.
\end{equation}
The Weyl scalars  for this rotating  solution are
\begin{eqnarray*}
\psi_2={1\over\overline R\,\overline R\,R^2}\Big[e^2
-R\,\{m+f(u)\}\Big]
\end{eqnarray*}
\begin{equation}
\psi_3 ={-i\,a\,{\rm sin}\theta\over 2\surd 2\overline
R\,\overline R\,R^2}
\Big\{4\,r\,f(u)_{,u}+\overline R\,f(u)_{,u}\Big\},
\end{equation}
\begin{eqnarray*}
\psi_4 ={{a^2r\,{\rm sin}^2\theta}\over 2\overline
R\,\overline R\,R^2\,R^2}\,\Big\{R^2f(u)_{,uu}-2rf(u)_{,u}\Big\}.
\end{eqnarray*}
This represents a rotating non-stationary Kerr-Newman-Vaidya
solution with the line element
\begin{eqnarray}
ds^2&=&[1-R^{-2}\{2r(m+f(u))-e^2\}]\,du^2+2du\,dr \cr
&&+2aR^{-2}\{2r(m+f(u))-e^2\}\,{\rm sin}^2
\theta\,du\,d\phi
-2a\,{\rm sin}^2\theta\,dr\,d\phi \cr
&&-R^2d\theta^2-\{(r^2+a^2)^2
-\Delta^*a^2\,{\rm sin}^2\theta\}\,R^{-2}{\rm sin}^2\theta\,d\phi^2,
\end{eqnarray}
where $\Delta^*=r^2-2r\{m+f(u)\}+a^2+e^2$. Here $m$ and $e$ are
the mass and the charge of Kerr-Newman solution, $a$ is the
rotation per unit mass and $f(u)$ represents the mass function of
rotating Vaidya null radiating fluid. The solution (6.21) will
describe a black hole if $m+f(u)>a^2+e^2$ with external event
horizon at $r_{+}= \{m+f(u)\}+\surd {[\{m+f(u)\}^2-a^2-e^2]}$, an
internal Cauchy horizon at $r_{-}=\{m+f(u)\} -\surd{[\{m+f(u)\}^2
-a^2-e^2]}$ and the non-stationary limit surface $r\equiv
r_e(u,\theta)= \{m+f(u)\} +\surd {[\{m+f(u)\}^2-a^2{\rm
cos}^2\theta-e^2]}$. When we set $f(u)=0$, the metric (6.21)
recovers the standard Kerr-Newman black hole, and if $m=0$, then
it is the rotating charged Vaidya null radiating black hole (5.9).

In this rotating solution, the Vaidya null fluid is interacting
with the non-null electromagnetic field whose Maxwell scalar
$\phi_1$ can be obtained from (6.18). Thus, we can write the total
energy momentum tensor (EMT) for the rotating solution (6.21) as
follows:
\begin{eqnarray}
T_{ab} = T^{(\rm n)}_{ab} +T^{(\rm E)}_{ab},
\end{eqnarray}
where the EMTs for the rotating null fluid as well as that of the
electromagnetic field are given respectively
\begin{eqnarray}
T^{(\rm n)}_{ab}= \mu^*\,\ell_a\,\ell_b +
2\,\omega\,\ell_{(a}\,\overline
m_{b)}+2\,\overline\omega\,\ell_{(a}\,m_{b)},
\end{eqnarray}
\begin{eqnarray}
T^{(\rm E)}_{ab}=2\,\rho^*\,\{\ell_{(a}\,n_{b)}
 + m_{(a}\,\overline m_{b)}\}.
\end{eqnarray}
The appearance of non-vanishing $\omega$ shows the null fluid is
rotating as the expression (6.19) of $\omega$ involves the
rotating constant $a$ coupling with $\partial f(u)/\partial u$ -
both are non-zero quantities for a rotating Vaidya null radiating
universe (5.9).

This Kerr-Newman-Vaidya metric (6.21) can be written in
Kerr-Schild ansatz on the Kerr-Newman background as
\begin{equation}
g_{ab}^{\rm KNV}=g_{ab}^{\rm KN} +2Q(u,r,\theta)\ell_a\ell_b
\end{equation}
where
\begin{equation}
Q(u,r,\theta) =-rf(u)R^{-2},
\end{equation}
and the vector $\ell_a$ is a geodesic, shear free, expanding as
well as rotating null vector of both $g_{ab}^{\rm KN}$ as well as
$g_{ab}^{\rm KNV}$ and given in (2.8) and $g_{ab}^{\rm KN}$ is the
Kerr-Newman metric (5.4) with $m=e$ = constant. This null vector
$\ell_a$ is one of the double repeated principal null vectors of
the Weyl tensor of $g_{ab}^{\rm KN}$. This completes the proof of
theorem 1 stated above.

   It appears from (6.25) that the Kerr-Newman geometry can be thought
of joining smoothly to the rotating Vaidya geometry at its null
radiative boundary, as shown by Glass and Krisch [23] in the case
of Schwarzschild geometry joining to the non-rotating Vaidya
space-time. The Kerr-Schild form (6.25) will recover that of
Xanthopoulos [33] $g'_{ab}=g_{ab}+\ell_a\ell_b$, when
$Q(u,r,\theta) \rightarrow 1/2$ and that of Glass and Krisch [23]
$g'_{ab}=g_{ab}^{\rm Sch}-\{2f(u)/r\}\ell_a\ell_b$ when $e=a=0$
for non-rotating Schwarzschild background space.
Thus, it can be regarded that the Kerr-Schild form presented in
(6.25) above will be the extension of those of Xanthopoulos as
well as Glass and Krisch. When we set $a=0$, this
Kerr-Newman-Vaidya solution (6.21) will recover to non-rotating
Reissner-Nordstrom-Vaidya solution with the Kerr-Schild form
$g_{ab}^{\rm RNV}=g_{ab}^{\rm RN} - \{2f(u)/r\}\ell_a\ell_b$,
which is still a generalization of Xanthopoulos and Glass and
Krisch in the charged Reissner-Nordstrom solution. It is worth to
mention that the new solution (6.21) cannot be considered as a
bimetric theory as $g_{ab}^{\rm KNV}\neq{1\over2}(g_{ab}^{\rm
KN}+g_{ab}^{\rm V})$.

To interpret the Kerr-Newman-Vaidya solution as a black hole
during the early inflationary phase of rotating Vaidya null
radiating universe {\it i.e.}, the Kerr-Newman black hole embedded
in rotating Vaidya null radiating background space, we can also
write the Kerr-Schild form (6.25) as
\begin{equation}
g_{ab}^{\rm KNV}=g_{ab}^{\rm V} +2Q(r,\theta)\ell_a\ell_b
\end{equation}
where
\begin{equation}
Q(r,\theta) =-(rm - e^2/2)R^{-2},
\end{equation}
Here, the constants $m$ and $e$ are the mass and the charge of
Kerr-Newman black hole, $g_{ab}^{\rm V}$ is the rotating Vaidya
null radiating black hole (5.9) and $\ell_a$ is the geodesic null
vector given in (2.8) for both $g_{ab}^{\rm KNV}$ and $g_{ab}^{\rm
V}$. When we set $f(u)=a=0$, $g_{ab}^{\rm V}$ will recover the
flat metric, then  $g_{ab}^{\rm KNV}$ becomes the original
Kerr-Schild form written in spherical symmetric flat background.

These two Kerr-Schild forms (6.25) and (6.27) certainly confirm
that the metric  $g_{ab}^{\rm KNV}$  is a solution of Einstein's
field equations since the background rotating metrics $g_{ab}^{\rm
KN}$ and  $g_{ab}^{\rm V}$ are solutions of Einstein's equations.
They both have different stress-energy tensors $T_{ab}^{(\rm E)}$
and $T_{ab}^{(\rm n)}$ given in (6.24) and (6.23) respectively.
Looking at the Kerr-Schild form (6.27), the Kerr-Newman-Vaidya
black hole can be treated as a generalization of Kerr-Newman black
hole by incorporating Visser's suggestion [34] that {\it
Kerr-Newman black hole embedded in an axisymmetric cloud of matter
would be of interest}. Hawking et al [35] have also mentioned the
possibility to embed the rotating black hole solutions with a
theory for which they know the corresponding conformal field
theory.

\subsubsection{\it Kerr-Newman-Vaidya-Bonnor
solution}

Similarly, one can combine the rotating Vaidya-Bonnor solution
obtained above in (5.13) with the Kerr-Newman solution (5.4) to
generate another rotating solution with the mass function
\begin{equation}
M(u,r)=m+f(u)-(e^2+h^2(u))/2r,
\end{equation}
representing a Kerr-Newman-Vaidya-Bonnor solution:
\begin{eqnarray}
ds^2&=&[1-R^{-2}\{2r(m+f(u))-e^2-h^2(u)\}]\,du^2+2du\,dr \cr
&&+2aR^{-2}\{2r(m+f(u))-e^2-h^2(u)\}\,{\rm sin}^2
\theta\,du\,d\phi
-2a\,{\rm sin}^2\theta\,dr\,d\phi \cr
&&-R^2d\theta^2-\{(r^2+a^2)^2
-\Delta^*a^2\,{\rm sin}^2\theta\}\,R^{-2}{\rm sin}^2\theta\,d\phi^2,
\end{eqnarray}
where $\Delta^*=r^2-2r\{m+f(u)\}+a^2+e^2+h^2(u)$. This rotating
solution can also be written in Kerr-Schild form (6.25) with the
function:
\begin{eqnarray*}
Q(u,r,\theta) =-\{r\,f(u)-h^2(u)/2\}\,R^{-2},
\end{eqnarray*}
When the charge $e$ of the Kerr-Newman solution vanishes, this
{\sl rotating} solution (6.30) will reduce to a rotating
Kerr-Vaidya-Bonnor solution with the mass function:
\begin{equation}
M(u,r)=m+f(u)-h^2(u)/2r.
\end{equation}

   It suggests that by choosing the Wang-Wu functions $q_n(u)$
properly one can generate as many rotating solutions as required.
However, the generation of these types of rotating solutions will
be restricted that  the energy-momentum tensor of the fluids must
be of the form given in (4.1).

\subsubsection{\it Kerr-Newman-de Sitter metrics}

Here we shall present the rotating de Sitter as well as
Kerr-Newman-de Sitter metrics in NP formalism.  \\\\
A. {\it  Rotating de Sitter solution}: \\\\
First we will derive a rotating de Sitter solution of Einstein's
equations. For this we choose the Wang-Wu functions as
\begin{eqnarray}
\begin{array}{cc}
q_n(u)=&\left\{\begin{array}{ll}
\Lambda^*/6, &{\rm when}\;\;n=3\\
0, &{\rm when }\;\;n\neq 3
\end{array}\right.
\end{array}
\end{eqnarray}
to obtain the mass function
\begin{equation}
M(u,r)={\Lambda^*\,r^3\over 6},
\end{equation}

The line element for the rotating de Sitter metric is
\begin{eqnarray}
ds^2&=&\Big\{1-{\Lambda^*\,r^4\over3\,R^2}
\Big\}\,du^2
+2du\,dr \cr
&&+2a{\Lambda^*\,r^4\over3}\,R^{-2}\,{\rm
sin}^2\theta\,du\,d\phi-2a\,{\rm sin}^2\theta\,dr\,d\phi \cr
&&-R^2d\theta^2-\Big\{(r^2+a^2)^2
-\Delta^*a^2\,{\rm sin}^2\theta\Big\}\,R^{-2}{\rm
sin}^2\theta\,d\phi^2,
\end{eqnarray}
where $R^2=r^2+a^2{\rm cos}^2\theta$,
$\Delta^*=r^2-{\Lambda^*\,r^4}/3+a^2$. This corresponds to the
rotating de Sitter solution for $\Lambda^*>0$, and to the anti-de
Sitter solution for $\Lambda^*<0$. In general $\Lambda^*$ denotes
the cosmological constant of the de Sitter space. Then the changed
NP quantities are
\begin{eqnarray*}
\gamma=-{1\over{2\overline R\,R^2}}\,\left\{(1-{1\over
3}\Lambda^*r^2)r\,\overline R+\Delta^*\right\},
\end{eqnarray*}
\begin{equation}
\phi_{11}= -{1\over {2\,R^2\,R^2}}\Lambda^*r^2a^2\,{\rm
cos}^2\theta,
\end{equation}
\begin{equation}
\psi_2={1\over{3\overline R\,\overline
R\,R^2}}\Lambda^*r^2a^2{\rm
cos}^2\theta,
\end{equation}
\begin{equation}
\Lambda={\Lambda^*r^2\over{6R^2}}.
\end{equation}
This means that in rotating de Sitter cosmological universe, the
$\Lambda^*$ is coupling with the rotational parameter $a$. From
these NP quantities we can clearly observe that the rotating de
Sitter cosmological metric is a Petrov type $D$ gravitational
field whose one of the repeated principal null vectors, $\ell_a$
is geodesic, shear free, expanding as well as non-zero twist. The
rotating cosmological space possesses an energy-momentum tensor
\begin{eqnarray}
T_{ab} =2\,\rho^*\,\ell_{(a}\,n_{b)}
+2\,p\,m_{(a}\overline m_{b)},
\end{eqnarray}
where $K\,\rho^*=2\,\phi_{11} + 6\,\Lambda$ and $K\,p=2\,\phi_{11}
- 6\,\Lambda$ are related to the density and the pressure  of the
cosmological matter which is, however not a perfect fluid. If we
set the rotational parameter $a=0$, we will recover the
non-rotating de Sitter metric [36], which is a solution of the
Einstein's equations for an empty space with $\Lambda \equiv
g_{ab}R^{ab}=\Lambda^*/6$ or constant curvature. However, it is
observed that the rotating de Sitter metric (6.34) is neither {\sl
empty} nor {\sl constant curvature}. It certainly describes a
stationary rotating spherical symmetric solution representing
Petrov type $D$ spacetime. So it is noted that to the best of the
present author's knowledge, this rotating de Sitter metric
has not been seen discussed  before. \\\\\
B. {\it Kerr-Newman-de Sitter solution}: \\\\\
By choosing the Wang-Wu function as
\begin{eqnarray}
\begin{array}{cc}
q_n(u)=&\left\{\begin{array}{ll}
m,&{\rm when }\;\;n=0\\
-e^2/2, &{\rm when }\;\;n=-1\\
\Lambda^*/6, &{\rm when}\;\;n=3\\
0, &{\rm when }\;\;n\neq 0, -1, 3
\end{array}\right.
\end{array}
\end{eqnarray}
we can obtain the function from (6.8)
\begin{equation}
M(u,r)=m-{e^2\over 2r}+{\Lambda^*\,r^3\over 6},
\end{equation}
where $m$ and $e$ are constants and  are the mass and the charge
of the Kerr-Newman solution. The line element with the function
(6.39) is
\begin{eqnarray}
ds^2&=&\Big\{1-R^{-2}\Big(2mr-e^2
+{\Lambda^*\,r^4\over3}\Big)
\Big\}\,du^2
+2du\,dr \cr
&&+2aR^{-2}\Big(2mr-e^2+{\Lambda^*\,r^4\over 3}\Big)\,{\rm
sin}^2\theta\,du\,d\phi-2a\,{\rm sin}^2\theta\,dr\,d\phi \cr
&&-R^2d\theta^2-\Big\{(r^2+a^2)^2
-\Delta^*a^2\,{\rm sin}^2\theta\Big\}\,R^{-2}{\rm
sin}^2\theta\,d\phi^2,
\end{eqnarray}
where $R^2=r^2+a^2{\rm cos}^2\theta$,
$\Delta^*=r^2-2mr-{\Lambda^*\,r^4}/3+a^2+e^2$.
Then the changed NP quantities are
\begin{eqnarray}
\gamma&=&{1\over{2\overline R\,R^2}}\,\left[\Big(r-m-{2\over
3}\Lambda^*r^3\Big)\overline R-\Delta^*\right],    \cr
\phi_{11}&=& {1\over {2\,R^2\,R^2}}\Big(e^2-\Lambda^*r^2a^2\,{\rm
cos}^2\theta\Big),                          \cr
\psi_2&=&{1\over{\overline R\,\overline R\,R^2}}\Big\{e^2-m\,R
+{\Lambda^*r^2\over 3}a^2{\rm cos}^2\theta\Big\}       \cr
\Lambda&=&{\Lambda^*r^2\over{6R^2}}.
\end{eqnarray}
We have seen from the above that in each expression of $\phi_{11}$
and $\psi_2$, the cosmological constant $\Lambda^*$ is coupling
with the rotational parameter $a$. This means that the
cosmological parameter $\Lambda^*$ has the effect of its presence
in the curvature of the embedded Kerr-Newman black hole. The
metric (6.41) admits the following energy momentum tensor
\begin{eqnarray*}
T_{ab} =2\,\rho^*\,\ell_{(a}\,n_{b)}
+2\,p\,m_{(a}\overline m_{b)},
\end{eqnarray*}
with the density and the pressure of the matter field
\begin{eqnarray*}
\rho^*&=& {1\over {K\,R^2\,R^2}}\Big(e^2+\Lambda^*r^4\Big),
\end{eqnarray*}
\begin{eqnarray*}
p&=& {1\over {K\,R^2\,R^2}}\Big\{e^2
-\Lambda^*r^2\Big(r^2+2a^2\,{\rm cos}^2\theta\Big)\Big\}
\end{eqnarray*}
respectively. Without loss of generality, we can write this
$T_{ab}$ with the decomposition of $\rho^*=\rho^{*(\rm
E)}+\rho^{*(\rm C)}$ and $p=p^{(\rm E)} + p^{(\rm C)}$ that
\begin{eqnarray}
T_{ab} =4\,\rho^{*(\rm E)}\{\ell_{(a}\,n_{b)}
+m_{(a}\overline m_{b)}\} +
2\{\rho^{*(\rm C)}\,\ell_{(a}\,n_{b)}
+2\,p^{(\rm C)}\,m_{(a}\overline m_{b)}\},
\end{eqnarray}
where
\begin{eqnarray*}
\rho^{*(\rm E)}&=&p^{(\rm E)}= {e^2\over {K\,R^2\,R^2}},\cr
\rho^{*(\rm C)}&=& {\Lambda^*r^4\over {K\,R^2\,R^2}}, \:\:\:\:
p^{(\rm C)}= {-\Lambda^*r^2\over{K\,R^2\,R^2}} \Big(r^2+2a^2\,{\rm
cos}^2\theta\Big).
\end{eqnarray*}
The advantage of writing $T_{ab}$ in the form (6.43) is that, for
Reissner-Nordstrom-de Sitter metric $(a=0)$, the energy-momentum
tensor can be written in the form of Guth's modification of
$T_{ab}$ for inflationary universe [37] as
\begin{equation}
T_{ab} =T^{(\rm E)}_{ab} + \Lambda^*g_{ab}
\end{equation}
where $T^{(\rm E)}_{ab}$ is the energy-momentum tensor for
non-null electromagnetic field and $g_{ab}$ is the
Reissner-Nordstrom-de Sitter metric. This indicates that Guth's
modification of $T_{ab}$ is acceptable only in the case of
non-rotating metrics, and its extension in the case of the
rotating solutions $(a \neq 0)$ will take the form given in (6.43)
above where $\rho^{*(\rm C)}\neq p^{(\rm C)}$.

The metric (6.41) describes a rotating stationary solution and is
Petrov type $D$ with $(\psi_2\neq 0)$, whose one of the repeated
principal null directions is $\ell_a$. That is, the metric can be
written in Kerr-Schild form  on the de Sitter background as
\begin{equation}
g_{ab}^{\rm KNdS}=g_{ab}^{\rm dS} +2Q(r,\theta)\ell_a\ell_b
\end{equation}
where $Q(r,\theta) =-(rm-e^2/2)R^{-2}$, and the vector $\ell_a$ is
a geodesic, shear free, expanding as well as rotating null vector
of both $g_{ab}^{\rm KNdS}$ as well as $g_{ab}^{\rm dS}$ and given
in (2.8) and $g_{ab}^{\rm KN}$ is the Kerr-Newman metric (5.4)
with $m=e$ = constant. This null vector $\ell_a$ is one of the
double repeated principal null vectors of the Weyl tensor of
$g_{ab}^{\rm KNdS}$ and $g_{ab}^{\rm dS}$. This completes the
proof of theorem 2 stated above. We can also write the Kerr-Schild
form (6.45) on the Kerr-Newman background as
\begin{eqnarray}
g_{ab}^{\rm KNdS}=g_{ab}^{\rm
KN}+2Q(r,\theta)\,\ell_a\,\ell_b,
\end{eqnarray}
where $Q(r,\theta)=-(\Lambda^*r^4/6)\,R^{-2}$, and $\Lambda^*$ is
the cosmological constant and $g_{ab}^{\rm KN}$ is the Kerr-Newman
metric (5.4) with $m=e$ = constant.

It is quite natural to recover rotating de Sitter
($m=0,a\neq0,e=0$) and Reissner-Nordstrom-de Sitter
($m\neq0,a=o,e\neq0$) metrics from the Kerr-Newman-de Sitter
solution ($m\neq0,a\neq0,e\neq0$). It is also worth mentioning
that from the Kerr-Newman-de Sitter metric, we can recover a
rotating charged de Sitter cosmological universe ($m=0,a\neq
0,e\neq0$). It is noted that one may find the difference between
this Kerr-Newman-de Sitter metric (6.41) and that of Mallett [8]
used by Koberlin [38]. Mallett's derivation of Kerr-Newman-de
Sitter metric is based on the direct application of Newman-Janis
algorithm to the Reissner-Nordstrom-de Sitter `seed' solution. It
is also found that the Kerr-Newman-de Sitter solution (6.41) is
different, in the terms involving the cosmological constant
$\Lambda^*$, from the one derived by Carter [39] and used by
Gibbon and Hawking [40], Khanal [41], Hawking, et al [35], and
others.
\subsubsection{\it Kerr-Newman-Vaidya-de Sitter
solution}

Here we shall combine the Kerr-Newman-de Sitter solution (6.41)
with the rotating Vaidya solution given in (5.9). For this purpose
we choose the Wang-Wu functions $q_n(u)$ as follows
\begin{eqnarray}
\begin{array}{cc}
q_n(u)=&\left\{\begin{array}{ll}
m+f(u),&{\rm when }\;\;n=0\\
-e^2/2, &{\rm when }\;\;n=-1\\
\Lambda^*/6, &{\rm when}\;\;n=3\\
0, &{\rm when }\;\;n\neq 0, -1, 3,
\end{array}\right.
\end{array}
\end{eqnarray}
where $m$ and $e$ are constants and $f(u)$ is related with the
mass of rotating Vaidya solution (5.9). Thus, using (6.47) in
(6.8) we have the mass function as
\begin{eqnarray*}
M(u,r)=m+f(u)-{e^2\over 2r}+{\Lambda^*\,r^3\over 6}
\end{eqnarray*}
and other quantities are
\begin{eqnarray*}
\rho^* &=& {1\over K\,R^2\,R^2}(e^2 + \Lambda^*r^4),
\end{eqnarray*}
\begin{eqnarray}
p&=& {1\over K\,R^2\,R^2}\Big\{e^2 - \Lambda^*r^2(r^2 +2a^2{\rm
cos}\theta)\Big\},
\end{eqnarray}
\begin{eqnarray}
\mu^* &=&-{r\over K\,R^2\,R^2}\Big\{2\,r\,f(u)_{,u}+a^2{\rm
sin}^2\theta\,f(u)_{,uu}\Big\},                           \cr
\omega &=&{-i\,a\,{\rm sin}\,\theta\,\over{\surd 2\,K\,\overline
R\,R^2}}\,f(u)_{,u},                                       \cr
\Lambda&\equiv & g^{ab}\,R_{ab}={\Lambda^*r^2 \over 6\,R^2},
\end{eqnarray}
and $\phi_{11}$, $\phi_{12}$, $\phi_{22}$ can be obtained from
equations (6.48) with (3.10). The Weyl scalars are given below
\begin{eqnarray}
\psi_2&=&{1\over\overline R\,\overline R\,R^2}\Big[e^2
-R\,\{m+f(u)\}+{\Lambda^*r^2\over 3}a^2{\rm cos}^2\theta\Big],\cr
\psi_3 &=&{-i\,a\,{\rm sin}\theta\over 2\surd 2\overline
R\,\overline R\,R^2} \Big\{\Big(4\,r+\overline
R\Big)f(u)_{,u}\Big\}, \cr
\psi_4 &=&{{a^2r\,{\rm
sin}^2\theta}\over 2\overline R\,\overline
R\,R^2\,R^2}\,\Big\{R^2f(u)_{,uu}-2rf(u)_{,u}\Big\}.
\end{eqnarray}
This represents a Kerr-Newman-Vaidya-de Sitter solution with the
line element
\begin{eqnarray}
d s^2&=&\Big[1-R^{-2}\Big\{2r\Big(m+f(u)\Big)+{\Lambda^*\,r^4\over
3}-e^2\Big\}\Big]\,du^2+2du\,dr \cr
&&+2aR^{-2}\Big\{2r\Big(m+f(u)\Big)+{\Lambda^*\,r^4\over 3}
-e^2\Big\}\,{\rm sin}^2 \theta\,du\,d\phi -2a\,{\rm
sin}^2\theta\,dr\,d\phi \cr &&-R^2d\theta^2 -\Big\{(r^2+a^2)^2
-\Delta^*a^2\,{\rm sin}^2\theta\Big\}\,R^{-2}{\rm
sin}^2\theta\,d\phi^2,
\end{eqnarray}
where $\Delta^*=r^2-2r\{m+f(u)\}-\Lambda^*\,r^4/3+a^2+e^2$. Here
$m$ and $e$ are the mass and the charge of Kerr-Newman solution,
$a$ is the  non-zero rotation parameter per unit mass and $f(u)$
represents the mass function of rotating Vaidya null radiating
fluid. When we set $f(u)=0$, the metric (6.51) recovers the
Kerr-Newman-de Sitter black hole (6.41), and if $m=0$, then it is
the rotating charged Vaidya null radiating black hole (5.13). When
one sets $\Lambda^*=0$, this metric will recover the
Kerr-Newman-Vaidya metric ((6.21). In this rotating solution, the
Vaidya null fluid is interacting with the non-null electromagnetic
field whose Maxwell scalar $\phi_1$ can be obtained from (6.48).
Thus, we can write the total energy momentum tensor (EMT) for the
rotating solution (6.51) as follows:
\begin{eqnarray}
T_{ab} = T^{(\rm n)}_{ab} +T^{(\rm E)}_{ab}+T^{(\rm C)}_{ab},
\end{eqnarray}
where the EMTs for the rotating null fluid, the electromagnetic
field and cosmological matter field are given respectively
\begin{eqnarray}
T^{(\rm n)}_{ab}&=& \mu^*\,\ell_a\,\ell_b +
2\,\omega\,\ell_{(a}\,\overline
m_{b)}+2\,\overline\omega\,\ell_{(a}\,m_{b)}, \cr
T^{(\rm E)}_{ab}
&=&4\,\rho^{*(\rm E)}\{\ell_{(a}\,n_{b)} +m_{(a}\overline
m_{b)}\}, \cr T^{(\rm C)}_{ab}&=&2\{\rho^{*(\rm
C)}\,\ell_{(a}\,n_{b)} +2\,p^{(\rm C)}\,m_{(a}\overline m_{b)}\},
\end{eqnarray}
where $\mu^*$ and $\omega$ are given in (6.49) and
\begin{eqnarray}
\rho^{*(\rm E)}&=&p^{(\rm E)}= {e^2\over {K\,R^2\,R^2}}, \cr
\rho^{*(\rm C)}&=& {\Lambda^*r^4\over {K\,R^2\,R^2}}, \:\:\:\:
p^{(\rm C)}= -{\Lambda^*r^2\over{K\,R^2\,R^2}} \Big(r^2+2a^2\,{\rm
cos}^2\theta\Big).
\end{eqnarray}
The appearance of non-vanishing $\omega$ shows that the null fluid
is rotating as the expression of $\omega$ (6.49) involves the
rotating parameter $a$ coupling with $\partial f(u)/\partial u$ --
both are non-zero quantities for a rotating Vaidya null radiating
universe.

This Kerr-Newman-Vaidya-de Sitter metric (6.51) can be written in
Kerr-Schild form on the de Sitter background as
\begin{equation}
g_{ab}^{\rm KNVdS}=g_{ab}^{\rm dS} +2Q(u,r,\theta)\ell_a\ell_b
\end{equation}
where
\begin{equation}
Q(u,r,\theta) =-[r\{m+f(u)\}-e^2/2]R^{-2},
\end{equation}
and the vector $\ell_a$ is a geodesic, shear free, expanding as
well as rotating null vector of both $g_{ab}^{\rm dS}$ as well as
$g_{ab}^{\rm KNVdS}$ and given in (2.8). We can also write this
solution (6.51) in another Kerr-Schild form on the Kerr-Newman
background as
\begin{equation}
g_{ab}^{\rm KNVdS}=g_{ab}^{\rm KN} +2Q(u,r,\theta)\ell_a\ell_b
\end{equation}
where $Q(u,r,\theta) =-\{rf(u)+\Lambda^*r^4/6\}R^{-2}$. These two
Kerr-Schild forms (6.55) and (6.57) certainly confirm that the
metric $g_{ab}^{\rm KNVdS}$  is a solution of Einstein's field
equations since the background rotating metrics  $g_{ab}^{\rm KN}$
and $g_{ab}^{\rm dS}$ are both solutions of Einstein's field
equations. They both have different stress-energy tensors
$T_{ab}^{(\rm E)}$ and $T_{ab}^{(\rm C)}$ given in (6.53).

\subsubsection{\it Rotating Vaidya-Bonnor-de Sitter
solution}

We shall combine the rotating Vaidya-Bonnor solution (5.13) with
the rotating de Sitter solution obtained above in (6.34), if the
Wang-Wu functions $q_n(u)$ are chosen such that
\begin{eqnarray}
\begin{array}{cc}
q_n(u)=&\left\{\begin{array}{ll}
f(u),&{\rm when }\;\;n=0\\
-e^2(u)/2, &{\rm when }\;\;n=-1\\
\Lambda^*/6, &{\rm when}\;\;n=3\\
0, &{\rm when }\;\;n\neq 0, -1, 3,
\end{array}\right.
\end{array}
\end{eqnarray}
where $f(u)$ and $e(u)$ are related with the mass and the charge
of rotating Vaidya-Bonnor solution (5.13). Thus, using this
$q_n(u)$ in (6.8) we obtain the mass function
\begin{equation}
M(u,r)=f(u)-{e^2(u)\over 2r}+{\Lambda^*\,r^3\over 6},
\end{equation}
and other quantities are
\begin{eqnarray*}
\rho^* &=& {1\over K\,R^2\,R^2}\Big\{e^2(u) + \Lambda^*r^4\Big\},
\end{eqnarray*}
\begin{eqnarray}
 p &=& {1\over K\,R^2\,R^2}\Big\{e^2(u) -
\Lambda^*r^2(r^2 +2a^2{\rm cos}\theta)\Big\},
\end{eqnarray}
\begin{eqnarray*}
\mu^* &=&-{1\over K\,R^2\,R^2}\Big[2\,r^2\Big\{f(u)_{,u}-{1\over
r}e(u)\,e(u)_{,u}\Big\}+a^2{\rm
sin}^2\theta\,\Big\{f(u)_{,u}-{1\over
r}e(u)\,e(u)_{,u}\Big\}_{,u}\Big],
\end{eqnarray*}
\begin{eqnarray*}
\omega &=&{-i\,a\,{\rm sin}\,\theta\,\over{\surd
2\,K\,R^2R^2}}\Big\{R\,f(u)_{,u}-2e(u)\,e(u)_{,u}\Big\}.
\end{eqnarray*}
and $\phi_{11}$, $\phi_{12}$, $\phi_{22}$ can be obtained from
equations (6.60) with (3.10). The Weyl scalars are given below
\begin{eqnarray}
\psi_2&=&{1\over\overline R\,\overline R\,R^2}\Big[e^2(u)
-R\,f(u)+{\Lambda^*r^2\over 3}a^2{\rm cos}^2\theta\Big], \cr
\psi_3 &=&{-i\,a\,{\rm sin}\theta\over 2\surd 2\overline
R\,\overline R\,R^2} \Big[\Big(4\,r+\overline
R\Big)f(u)_{,u}\Big\}- 4e(u)\,e(u)_{,u}\Big], \cr
\psi_4&=&{{a^2\,{\rm sin}^2\theta}\over 2\overline R\,\overline
R\,R^2\,R^2}\,\Big[R^2\Big\{r\,f(u)_{,u} -e(u)e(u)_{,u}\Big\}_{,u}
\cr &&-2r\Big\{r\,f(u)_{,u} -e(u)e(u)_{,u}\Big\}\Big].
\end{eqnarray}
This represents a rotating Vaidya-Bonnor-de Sitter solution with
the line element
\begin{eqnarray}
d
s^2&=&\Big[1-R^{-2}\Big\{2rf(u)+{\Lambda^*\,r^4\over
3}-e^2(u)\Big\}\Big]\,du^2+2du\,dr \cr
&&+2aR^{-2}\Big\{2rf(u)+{\Lambda^*\,r^4\over 3}
-e^2(u)\Big\}\,{\rm sin}^2
\theta\,du\,d\phi
-2a\,{\rm sin}^2\theta\,dr\,d\phi \cr
&&-R^2d\theta^2
-\Big\{(r^2+a^2)^2
-\Delta^*a^2\,{\rm sin}^2\theta\Big\}\,R^{-2}{\rm
sin}^2\theta\,d\phi^2,
\end{eqnarray}
where $\Delta^*=r^2-2rf(u)-{\Lambda^*\,r^3/3} +a^2+e^2(u)$.
Here, $a$ is the  non-zero rotational parameter per unit mass and
$f(u)$ represents the mass function of rotating Vaidya null
radiating fluid. We can write the total energy momentum tensor
(EMT) for the rotating solution (6.62) as follows:
\begin{eqnarray}
T_{ab} = T^{(\rm n)}_{ab} +T^{(\rm E)}_{ab}+T^{(\rm C)}_{ab},
\end{eqnarray}
where the EMTs for the rotating null fluid, the electromagnetic
field and cosmological matter field are given respectively
\begin{eqnarray}
T^{(\rm n)}_{ab}&=& \mu^*\,\ell_a\,\ell_b +
2\,\omega\,\ell_{(a}\,\overline
m_{b)}+2\,\overline\omega\,\ell_{(a}\,m_{b)},
\end{eqnarray}
\begin{eqnarray}
T^{(\rm E)}_{ab} &=&4\,\rho^{*(\rm E)}\{\ell_{(a}\,n_{b)}
+m_{(a}\overline m_{b)}\},
\end{eqnarray}
\begin{eqnarray}
T^{(\rm C)}_{ab}&=&2\{\rho^{*(\rm C)}\,\ell_{(a}\,n_{b)}
+2\,p^{(\rm C)}\,m_{(a}\overline m_{b)}\},
\end{eqnarray}
where
\begin{eqnarray*}
\rho^{*(\rm E)}&=& p^{(\rm E)}= {e^2(u)\over {K\,R^2\,R^2}}, \cr
\rho^{*(\rm C)}&=& {\Lambda^*r^4\over {K\,R^2\,R^2}}, \:\:\:\:
p^{(\rm C)}= -{\Lambda^*r^2\over{K\,R^2\,R^2}} \Big(r^2+2a^2\,{\rm
cos}^2\theta\Big).
\end{eqnarray*}
If we set $a=0$, we recover the non-rotating Vaidya-Bonnor-de
Sitter solution, and then the energy-momentum tensor (6.63) will
take the form of Guth's modification of $T_{ab}$ for inflationary
scenario [37] as
\begin{equation}
T_{ab} =T^{(\rm n)}_{ab}+T^{(\rm E)}_{ab} + \Lambda^*g_{ab}
\end{equation}
where $T^{(\rm E)}_{ab}$ is the energy-momentum tensor for
non-null electromagnetic field and $g_{ab}$ is the non-rotating
Vaidya-Bonnor-de Sitter metric tensor. From this, without loss of
generality, the EMT (6.63) can be regarded as the extension of
Guth's modification of energy-momentum tensor in rotating spaces.

The Vaidya-Bonnor-de Sitter metric can be written in Kerr-Schild
form
\begin{equation}
g_{ab}^{\rm VBdS}=g_{ab}^{\rm dS} +2Q(u,r,\theta)\ell_a\ell_b
\end{equation}
where $Q(u,r,\theta) = -\{rf(u)-e^2(u)/2\}\,R^{-2}$. Here,
$g_{ab}^{\rm dS}$ is the rotating de Sitter metric (6.34) and
$\ell_a$ is geodesic, shear free, expanding and non-zero twist
null vector for both $g_{ab}^{\rm dS}$ as well as $g_{ab}^{\rm
VBdS}$ and given in (2.8). The above Kerr-Schild form can also be
written on the rotating Vaidya-Bonnor background given in (5.13).
\begin{equation}
g_{ab}^{\rm VBdS}=g_{ab}^{\rm VB} +2Q(r,\theta)\ell_a\ell_b
\end{equation}
where $Q(r,\theta) =-(\Lambda^*r^4/6)R^{-2}$. These two
Kerr-Schild forms (6.68) and (6.69) prove the non-stationary
version of theorem 2 in the case of rotating Vaidya-Bonnor-de
Sitter solution. If we set $f(u)$ and $e(u)$ are both constant,
this Kerr-Schild form (6.68) will be that of Kerr-Newman-de Sitter
black hole (6.45). The rotating Vaidya-Bonnor-de Sitter metric
will describe a non-stationary spherically symmetric solution
whose Weyl curvature tensor is algebraically special in Petrov
classification possessing a geodesic, shear free, expanding and
non-zero twist null vector $\ell_a$ given in (2.8). One can easily
recover a rotating Vaidya-de Sitter metric from this
Vaidya-Bonnor-de Sitter solution by setting the charge $e(u)=0$.
If one sets $a=0$, $e(u)=0$ in (6.62), one can also obtain the
standard non-rotating Vaidya-de Sitter solution [42]. Ghosh and
Dadhich [43] have studied the gravitational collapse problem in
non-rotating Vaidya-de Sitter space by identifying the de Sitter
cosmological constant $\Lambda^*$ with the bag constant of the
null strange quark fluid. Also if one sets $a=0$ in (6.62) one can
recover the non-rotating Vaidya-Bonnor-de Sitter black hole [44].
It certainly indicates that all embedded solutions (6.21), (6.30),
(6.41), (6.62) can be derived by using Wang-Wu functions (6.8) in
the rotating solutions (6.4).

\section{Conclusion}
\setcounter{equation}{0}
\renewcommand{\theequation}{7.\arabic{equation}}

In this paper, we have calculated NP quantities for a rotating
spherically symmetric metric with three variables. With the help
of these NP quantities, we have first given examples of rotating
solutions like Kerr-Newman, rotating Vaidya and rotating
Vaidya-Bonnor. Then, with the help of Wang-Wu functions, we come
to the unpublished rotating metrics that we combined them with
other rotating solutions in order to derive new embedded rotating
solutions, and studied the gravitational structure of the
solutions by observing the nature of the energy-momentum tensors
of respective spacetime metrics. The embedded rotating solutions
have also been expressed in terms of Kerr-Schild forms in order to
show them as solutions of Einstein's field equations.

We would like to mention that Chandrasekhar [20] has established a
relation of spin coefficients $\rho$, $\mu$, $\tau$, $\pi$ in the
case of an affinely parameterized geodesic vector, generating an
integral which is constant along the geodesic in a vacuum Petrov
type $D$ space-time
\begin{equation}
{\rho\over \overline\rho}={\mu\over\overline\mu}={\tau\over\overline\pi}
={\pi\over\overline\tau}.
\end{equation}
This relation is being derived on the basis of the vacuum Petrov
type $D$ space-time with $\psi_2\neq 0$,
$\psi_0=\psi_1=\psi_3=\psi_4=0$ and $\phi_{01}=\phi_{02}=\phi_{10}
=\phi_{20}=\phi_{12}=\phi_{21}=\phi_{00}=\phi_{22}=\phi_{11}=\Lambda=0$.
However, it has been shown in [22] that the {\it non-vacuum}
Petrov type $D$ spacetimes {\it i.e.} Kerr-Newman solution
possessing electromagnetic field and Kantowski-Sachs metric with
dust energy-momentum tensor, satisfy the Chandrasekhar's relation
(7.1). We will show that  the very general metric (2.7) with three
variables, which is of {\it algebraically special} in Petrov
classification having non-zero $\psi_2$, $\psi_3$, $\psi_4$ and
stress-energy tensor (4.1), still satisfies the relation (7.1) as
follows
\begin{equation}
{\rho\over \overline\rho}={\mu\over\overline\mu}={\tau\over\overline\pi}
={\pi\over\overline\tau}={R\over\overline R}.
\end{equation}
This relation (7.2) shows that all rotating solutions, stationary
and non-stationary, discussed here satisfy the relation (7.1).
Thus, it seems reasonable to refer to the relation (7.1) as {\it
Chandrasekhar's identity} as mentioned by Fermandes and Lun [45].
This certainly indicates that the NP spin coefficients (3.3) can
be used to extend the known vacuum results like the relation (7.1)
to the non-vacuum ones. Further, we also observe from the
energy-momentum tensor (4.1) that the metric (2.7) with three
variables does not include the perfect fluid. From the above
discussion, it certainly indicates that the Newman-Janis algorithm
can be used to generate rotating solutions as shown in section 6,
if the metric function $M(u,r)$ is expressed in terms of Wang-Wu
functions given in (6.8). However such generated rotating
solutions from the application of Newman-Janis algorithm have
limitations that these rotating solutions are not included
spacetimes admitting rotating perfect fluid. To have a spacetime
admitting a rotating perfect fluid one has to look for another
algorithm rather than that of Newman and Janis.

From the above results presented in this paper, it suggests that
rotating spacetime geometries must also have rotating matter
fields, described by the stress-energy tensors $T_{ab}$ with non
zero rotation parameter $a$. It means that the matching of a
rotating non-stationary space-time geometry with the usual
non-rotating perfect fluid will not have a reasonably good sense.
So one needs to look for a rotating perfect fluid to match with a
rotating non-stationary spacetime geometry. Some of the rotating
solutions discussed above include rotating non-stationary
solutions, like Kerr-Newman-Vaidya black hole,
Kerr-Newman-Vaidya-de Sitter black hole, Vaidya-Bonnor-de Sitter
black hole. They possess rotating non-perfect fluids as shown
above by the respective $T_{ab}$. To study the nature of these
rotating black holes will certainly be a new area of interest in
classical General Relativity, since all known black hole theorems,
like `no hair theorem', Penrose's theorems are based on stationary
black holes, rotating or non-rotating.

It is also found that the technique of Wang and Wu with the
functions $q_n(u)$ in (6.8) can be used to generate rotating
solutions as shown in section 6.2 above. By choosing a suitable
Wang-Wu function $q(u)$, we obtain a rotating de Sitter space-time
model. We can also recover the widely used (i) Schwarzschild-de
Sitter solution, (ii) Reissner-Nordstrom-de Sitter black hole
solution, (iii) Kerr-de Sitter solution, (iv) Kerr-Newman-de
Sitter solution for early inflation scenarios from the rotating
Vaidya-Bonnor-de Sitter solution (6.62). These embedded de Sitter
space-times can generate by using Wang-Wu functions in rotating
solutions given in (6.9). This shows that Wang-Wu functions in
rotating space-time geometry can be applied to generate Kerr-de
Sitter and Kerr-Newman-de Sitter solutions in a simple way -- not
directly applying Newman-Janis algorithm to Schwarzschild-de
Sitter as well as Reissner-Nordstrom-de Sitter `seed' solutions to
derive Kerr-de Sitter as well as Kerr-Newman-de Sitter solutions
respectively. Thus, the rotating solutions (6.9) with Wang-Wu
functions can avoid the difficulties suggested by Xu [9]. The
definitions of embedded spaces used here are in agreement with the
one defined by Cai et al. [46]. It is worth mentioning that the
rotating embedded solution, namely Kerr-Newman-de Sitter solution
(6.41) is found different from the ones discussed in [8,9,40].
Hence, to the best of the author's knowledge, the rotating
embedded solutions (6.41), (6.51), (6.62) and other reducible
solutions from (6.62), and also (6.21), (6.30) have not been seen
published before. Other non-rotating embedded solutions of
Einstein's equations can be found in Kramer, et al. [47] (and
references there in) and Hodgkinson [48].

Looking at these overall rotating solutions derived above one can
conclude that
\begin{enumerate}
\item all stationary rotating solutions including (a) Kerr-Newman,
(b) rotating monopole, (c) rotating de Sitter, (d) Kerr-Newman de
Sitter solutions which are derivable from the application of
Newman-Janis algorithm, are Petrov type $D$ and each spacetime has
the repeated principal null vector $\ell_a$, which is geodesic,
shear free, expanding as well as rotating. This completes the
proof of theorem 3.

\item rotating Vaidya (5.9), rotating Vaidya-Bonnor (5.13),
Kerr-Newman-Vaidya (6.21), Kerr-Newman-Vaidya-Bonnor (6.26),
rotating Vaidya-de Sitter (6.62) when $e(u)=0$ and rotating
Vaidya-Bonnor-de Sitter (6.62) are all non-stationary spherically
symmetric solutions. Their Weyl curvature tensors are
algebraically special in the Petrov classification with null
vector $\ell_a$ given in (2.8). This leads the proof of the
theorem 4 stated in the introduction.
\end{enumerate}
The remarkable feature of the analysis  of rotating solutions in
this paper is that all the rotating solutions, stationary Petrov
type $D$ and non-stationary algebraically special, presented here
possess the same null vector $\ell_a$, which is geodesic, shear
free, expanding as well as non-zero twist. From the studies of the
rotating solutions we found that some solutions after making
rotation have disturbed their gravitational structure. For
example, the rotating monopole solution (6.11) possesses the
energy-momentum tensor with the monopole pressure $p$, where the
monopole constant $b$ couples with the rotating parameter $a$ .
Similarly, the rotating de Sitter solution (6.34) becomes Petrov
type $D$ spacetime metric, where the rotating parameter $a$ is
coupled with the cosmological constant and so on. We have shown
that all the rotating embedded solutions presented here can be
written in Kerr-Schild forms, showing the extension of those of
Xanthopoulos [33] and of Glass and Krisch [23].

\section*{Acknowledgement}

The author acknowledges his appreciation for hospitality received
from Inter-University Centre for Astronomy and Astrophysics
(IUCAA), Pune during the preparation of this paper.

\end{document}